\documentclass[lettersize,IEEE TRANSACTIONS ON SMART GRID]{IEEEtran}
\usepackage[english]{babel}
\usepackage{amsmath}
\usepackage{verbatim}
\usepackage{algorithm}
\usepackage{titlesec}
\PassOptionsToPackage{}{algpseudocode}
\usepackage{algpseudocode}%
\makeatletter
\def\BState{\State\hskip-\ALG@thistlm}
\makeatother
\algnewcommand\algorithmicinput{\textbf{Input:}}
\algnewcommand\algorithmicoutput{\textbf{Output:}}
\algnewcommand\Input{\item[\algorithmicinput]}%
\algnewcommand\Output{\item[\algorithmicoutput]}%
\usepackage{multicol}
\usepackage{multirow}
\usepackage{amssymb}
\usepackage{tikz}
\usepackage{booktabs}
\usepackage{array}
\usepackage[caption=false,font=normalsize,labelfont=sf,textfont=sf]{subfig}
\usepackage{textcomp}
\usepackage{float}
\usepackage{stfloats}
\usepackage{color}
\usepackage{xcolor}
\usepackage{soul}
\usepackage{url}
\usepackage{verbatim}
\usepackage{graphicx}
\usepackage{cite}
\usepackage{amsmath,amsthm,amsfonts}
\theoremstyle{definition}
\newtheorem{definition}{Definition}
\newtheorem{theorem}{Theorem}
\newtheorem{lemma}{Lemma}
\begin{document}

\title{SePEnTra: A secure and privacy-preserving energy trading mechanisms in transactive energy market}

\author{Rumpa Dasgupta, Amin Sakzad,  Carsten Rudolph,~\IEEEmembership{Member,~IEEE,} Rafael Dowsley \vspace{-1cm}}

\maketitle

\begin{sloppypar}
\begin{abstract}
Transactive energy market (TEM) has been emerging to balance dynamic demand-supply across the grid due to the large integration of distributed energy resources (DERs) in the electricity markets. TEM aims to bring fair compensation for all market participants, which heavily relies on accurate data originating from and propagating through different market components. Hence, maintaining data security with preserving individual’s privacy is essential as energy data is sensitive. In this paper, we design and present a novel model called SePEnTra to ensure the security and privacy of energy data while sharing with other entities during energy trading to determine optimal price signals. Furthermore, the market operator can store and use this data to detect malicious activities (deviation of actual energy generation/consumption from forecast data beyond a threshold) of users in the later stage without violating privacy. We use two cryptographic primitives, additive secret sharing and Pedersen commitment, in SePEnTra. We analyze the security of our proposed model in detail with proper security proofs. The performance of our proposed model is evaluated theoretically and numerically, considering practical TEM scenarios. We compare the performance of our model with the same TEM framework without security mechanisms. The result shows that even though using two advanced cryptographic primitives in a large market framework, our proposed model has very low computational complexity and communication overhead. Moreover, our model is storage efficient for both the market operator and users.

\end{abstract}


\vspace{-0.7cm}

\section{Introduction}
\subsection{Motivation}

\IEEEPARstart{T}{he} integration of distributed energy resources (DERs) in the electrical grid and their agile participation as a source of energy plays a crucial role in the modernization of the power systems. Moreover, through their engagement, DER owners convert from ordinary customers to active players in the energy market and demand response (DR) programs~\cite{Paul2015}. However, the intermittent nature of DER’s energy generation brings new challenges for grid operators and DER owners, deprived of their proper incentives from the conventional energy market. Hence, a new market framework called the transactive energy market (TEM)~\cite{Forfia2016a} has recently been designed for efficient energy management and fairly compensates all market participants. In TEM, DER owners can directly participate in the market operation without the involvement of utility companies. They can interact with each other directly or via some central market authority to sell or buy their surplus or deficit energy through active negotiation \cite{Omid2019}. TEM has three types of market structure, full peer-to-peer (P2P), community-based and hybrid market \cite{MKhorasany2020}. TEM uses auction-based or distributed market clearing process to settle the market and determine energy prices. In both approaches, energy demand and supply forecasts from market participants play a vital role in determining the fair energy price for all market participants.

Due to the rapid advance of information technology, market participants’ energy demand and supply forecasts are automatically generated from their home energy management system (HEMS) ~\cite{Dayaratne2023}. HEMS uses digital technologies to control, monitor, and receive energy data from all smart household appliances~\cite{Anon2015}. Any malicious and false information can mislead market operators in estimating market prices and balancing the energy supply and demand for the upcoming period. Intruders can take this opportunity and exploit other participants’ devices, such as smart appliances, DERs, HEMSs, etc., to generate and inject false predictions. Besides third-party attackers, some users may misbehave deliberately during the forecast phase to manipulate energy prices and obtain illegitimate economic benefits. Such malicious behaviour shifts the system equilibrium from the optimal solution and disrupts the regular market operation. Hence, an efficient technique is needed to detect malign activities of market participants.

Energy forecast is not only crucial for market authorities but also sensitive for market participants as user’s privacy is related to such information. Energy data gives an idea of user’s day-to-day activities, particularly in the daytime and at weekends. It can disclose critical information to an unauthorized person, such as a user's presence in the home, usage patterns of different electrical equipment, and personal preferences that violate user privacy \cite{Gong2016}. Therefore, measures need to be taken to preserve users' privacy while sharing energy consumption or supply forecast with the energy market.

\subsection{Related Works}
In the literature, several researchers worked on cyber security threats, vulnerabilities, attack detection, and mitigation mechanisms of TEM-based power systems. For instance, Zhang et al. ~\cite{Zhang2020} studied cyber-attacks based on threats in the transactive energy system (TES) and designed attack detection methods based on data analytics and machine learning. Note that some authors refer to TEM as TES; however, the functionalities of TEM and TES are identical. In~\cite{Jhala2019}, authors investigated the impact of false data injection attack (FDIA) on real-time electricity price, demand, and distribution system voltage. Barreto et al. ~\cite{Barreto2020} analyzed potential cyber risks for TEM caused by the internet of things integrated smart appliances and proposed a defence technique to mitigate the attack's impact. For several cyber attacks simulation, a Transactive Energy Security Simulation Testbed (TESST) was developed in~\cite{Zhang2019} by extending the transactive energy simulation platform (TESP) designed by the Pacific Northwest National Lab. However, these existing works~\cite{Jhala2019, Barreto2020, Zhang2019} mainly focused on attacks through communication links to change an individual’s bid during the bidding process or manipulate measurement data from smart meters. They overlooked the critical fact that false energy forecasts can be straight injected into the energy market from user’s control devices, such as HEMS or smart appliances, without manipulating communication. Hence, one possible way to detect the presence of false energy forecasts injected into the system by malicious users is to keep track of their prediction data. The main drawback of such a scheme is user privacy, as market operators have to preserve each user's energy data.  

Recently, considering the importance of user’s privacy, a few studies~\cite{Abdella2022,Aron2017,Kvaternik2017,Ullah2021,Lu2020,Yang2020} introduced privacy-preserved techniques in TEM. Table \ref{table:comparison} summarizes the market type, pricing mechanism, introduction of the third party besides regular market entities, FDIA through communication links and malicious users, attack detection, adopted techniques, and privacy considerations taken into account by these schemes. To be specific, Abdella et al.~\cite{Abdella2022} studied blockchain-based P2P energy trading systems. Laszka et al.~\cite{Aron2017} introduced a distributed-ledger-based solution called Privacy-preserving Energy Transactions (PETra)  during energy trading in transactive IoT microgrids to preserve prosumer privacy. Kvaternik et al.~\cite{Kvaternik2017} extended the above work~\cite{Aron2017} and implemented PETra using a microgrid's original load profile data set. However, they introduced blockchain technology with anonymity to preserve user privacy. Blockchain technology is not the perfect solution for TEM. The reason is if privacy is introduced into the blockchain, users may lose accountability in such cases. As a result, the system no more guarantees transparency, fraud identification, and detection. Hence, such a technology may not be useful in this scenario. A privacy-preserved distributed energy pricing scheme for the full-P2P TEM was developed in \cite{Ullah2021}. However, instead of focusing on attack or attacker detection, the primary objective of this work is to incorporate physical network constraints into energy pricing, specifically in regards to voltage and line congestion management.
Lu et al. developed a homomorphic encryption-based approach to preserve privacy during the market operation of TES in \cite{Lu2020} and \cite{Yang2020}. However, their schemes are unsuitable for real-time market operation due to the high computational cost \cite{Lu2020} and introduction of an honest third party (besides regular market entities such as market participants, transactive market operators, and distributed system operators (DSO) ) to perform encrypted aggregations~\cite{Yang2020}. Moreover, no work considered community-based TEM with distributed pricing mechanism as a system model to trade energy.
Therefore, taking the importance of the energy prediction information and user’s privacy into account, this work proposes a novel framework, which we call SePEnTra, for providing a secure and privacy-preserved energy trading mechanism in community-based TEM with distributed pricing mechanism.

\begin{table*}
\caption{Comparison of existing schemes with SePEnTra.}
\label{table:comparison}
\centering

\begin{tabular}{cccccccccccl}
\toprule
\multirow{2}{*}{\textbf{Ref.}}&\multicolumn{3}{c}{\textbf{Model}}&\multirow{1}{*}{}&
\multicolumn{2}{c}{\textbf{FDIA through}}&\textbf{Attack}&\multirow{2}{*}{\textbf{Technique}}&\multicolumn{2}{c}{\textbf{Privacy against}}&\textbf{Comput.}\\
\cline{2-4}\cline{6-7}\cline{10-11} 
&\textbf{Market}&\textbf{Pricing}&\textbf{3rd-party}&&\textbf{Links}&\textbf{Malicious} &\textbf{Detection}&&\textbf{Users}&\textbf{TO/DSO}&\textbf{Cost}\\
\midrule

\cite{Aron2017,Kvaternik2017}  & Full-P2P &Auction &\checkmark &  & X & X & X & Blockchain &\checkmark &X &11.79 s  \\
\cite{Ullah2021} & Full-P2P &Distributed &X &   & X & X &X &F-ADMM &\checkmark &X &1 s \\

\cite{Lu2020}  & Community & Auction &\checkmark & & X & X & X &Paillier enc. &\checkmark &\checkmark &22.31 s \\

\cite{Yang2020} & Community &Auction &\checkmark &   & \checkmark & X &\checkmark &Paillier enc. \& sig. &\checkmark &\checkmark &0.23 s \\

SePEnTra & Community &Distributed &X &  & \checkmark &\checkmark &\checkmark & Secret sharing \& Com.  &\checkmark &\checkmark &0.62 s  \\

\bottomrule
\multicolumn{12}{p{17.5cm}}{Notations: TO: Transactive Market Operator, DSO: Distributed System Operator, F-ADMM :  Fast Alternating Direction Method of Multipliers
}
\end{tabular}
\vspace{-0.6cm}
\end{table*}

\subsection{Contributions}
In this work, we present SePEnTra, a Secure Multiparty Computation (MPC) based novel framework for trading energy in TEM. More specifically, we use an additive secret sharing scheme and Pedersen commitment during energy trading. These primitives allow market participants to share energy consumption and generation forecast in such a way that participants' information remains private. However, market authorities/operators can use this information for energy price calculation in the early stage and malicious user’s detection in the later stage. To achieve this goal, we design a five-phase algorithm that consists of negotiation, one-off key generation, commitment, commitment check, and online phases. We analyze the security of our proposed model. We prove that except user him/herself, no other users or market operator learn anything about user’s energy demand/supply forecast and actual supply/consumption information. We also prove that our proposed model keeps data secure (not disclosed or view original information) from third-party attackers or unauthorized users during transmission from one entity (user) to another entity (market operator). Our model prevents manipulation of users’ energy demand or supply prediction commitment by third-party attackers, malicious users, and user him/herself. Using our model, we can detect malicious users whose actual energy generation/consumption deviates from energy forecasts beyond a threshold. We theoretically and numerically evaluate the computational costs, storage size, and communication costs of our SePEnTra scheme. We compare these results with the same market framework without adopting any security measures. Results show that our scheme's computational and communication cost is still low despite using advanced cryptographic primitives.

\section{Preliminaries}
\label{Preliminaries}

\subsection{Cryptographic Background}
In this section, we review some cryptographic primitives and their formal definitions that we opted for in our proposed model.

\subsubsection{Secure Multi-Party Computation (MPC)}
\label{MPC}
Secure multiparty computation (MPC) \cite{Lindell2020} is a well-known cryptographic technique that allows a number of distinct but connected computing devices or parties to jointly compute a function over their inputs securely. The essential requirement of any MPC protocol is privacy which states that parties should learn the output and nothing else to preserve participants’ privacy from each other. The privacy of a protocol is formally defined below:

\begin{definition}
[$t$-private protocol\cite{Desmedt2007}]
\label{def:tprivateprotocol}
Let $\left[ N \right]$ denote the set $\{ 1,\ldots, N \}$  where $N$ is the number of parties and  $F: (\{0,1\}^*)^N \rightarrow \{0,1\}^N$  be an $N$-bit input and single-output function. Let $\Pi$ denote an $N$-party protocol for computing $F$. The input sequence of each party is defined by $\boldsymbol{x} = (x_1,\ldots,x_N)$. The joint protocol view of parties in subset ${\mathcal{J}} \subseteq \left[ N \right]$ is denoted by  $\text {VIEW }_{\mathcal{J}}^{\Pi}(\boldsymbol{x})$, and the protocol output is defined by $\text { OUT }^{\Pi}(\boldsymbol{x})$. For $0 < t < N$, we say that $\Pi$ is a $t$-private protocol (secure against $t$ parties) for computing $F$ if there exists a probabilistic polynomial-time (PPT) algorithm $S$, such that, for every ${\mathcal{J}} \subset\left[ N \right]$ with $\# {\mathcal{J}}\leq t$ and every $\boldsymbol{x} \in (\{0,1\}^*)^n$, the random variables
\begin{equation*}
\left\langle S ({\mathcal{J}}, \boldsymbol{x}_{\mathcal{J}}, F(\boldsymbol{x})), F(\boldsymbol{x}) \right\rangle and \left\langle\text {VIEW }_{\mathcal{J}}^{\Pi}(\boldsymbol{x}), \text { OUT }^{\Pi}(\boldsymbol{x})\right\rangle
\end{equation*}
are identically distributed, where $\boldsymbol{x}_\mathcal{J}$ denotes the projection of the $N$-ary sequence $\boldsymbol{x}$ on the coordinates in $\mathcal{J}$ and $\# {\mathcal{J}}$ represents the size of set $\mathcal{J}$.

\end{definition}

\subsubsection{Additive Secret Sharing} 
\label{ASS}

Additive secret sharing is a secret-sharing-based MPC technique \cite{Daniel2022}, which we use in our solution. It divides secret data into several “shares” that sum up the original secret. While a secret is divided into a number of shares, each party holds a share. All shares need to be added to recover the original secret value. No proper subset of participants has enough information to reveal the secret. Formally, we can define additive secret sharing over a prime field $\mathbb{F}_p$ where $p$ is a prime number. 
For a secret sharing with $N$ participants, a secret $Y$ is shared by picking uniformly at random $N-1$ shares $[Y_1, Y_2, \ldots, Y_{N-1}] \in \mathbb{F}_p $. Then, $Y_N$ is set as $Y-\sum_{i=1}^{N-1} Y_i \in \mathbb{F}_p$. $Y$ can be reconstructed by adding $Y_1, \ldots, Y_N$. Any proper subset of shares gives no useful information to recover $Y$. 
Hence, additive secret sharing is a $(N,N)$ threshold secret sharing scheme, as  the secret is divided into $N$ shares, such that exactly $N$ shares allow the secret to being reconstructed, but less than $N$ shares reveal no information about the secret.

\begin{definition}
[Homomorphic Sharing Scheme \cite{Benaloh1987}]
\label{def:HSC1}
A secret sharing scheme is called homomorphic if the composition of the shares are shares of the composition. If $A$ is a secret defined by the shares $A_1,\ldots,A_N$ and $B$ is a secret defined by the shares $B_1,\ldots,B_N$,  then an additive secret sharing scheme is called homomorphic if 

\begin{equation*}
    A+B= A_1+\cdots+A_N+B_1+\cdots+B_N
\end{equation*}
\end{definition}

\begin{definition}
[Composite\cite{Desmedt93}]
\label{def:composite}
Let $U$ be the set secrets and $U=\{Y_1,\ldots,Y_N\}$. A homomorphic sharing scheme is composite in the strong sense if it is perfect and the revelation of all shadows of $Y_1 + \cdots + Y_N$, where $Y_1,\ldots Y_N \in U$, does not reveal anything more about $(Y_1,\ldots, Y_N)$ than what $Y_1 + \cdots + Y_N$ does.
\end{definition}

\subsubsection{Commitment Schemes}
\label{CS}
A commitment scheme \cite{Groth2015} is a cryptographic scheme that allows a committer or sender to commit to a chosen value. Later on, the sender can open the commitment and reveal the committed value. The recipient of the commitment can verify if the opened value is the same that was
committed in the earlier stage or not. A commitment scheme should be hiding and binding. Hiding means that the commitment does not reveal any information about the committed value after the first phase, and binding means that the committer cannot open the commitment to two different values. The formal definitions are given below.
\begin{definition}
[Commitment Scheme \cite{Groth2015}]
\label{def:cs}
A commitment scheme consists of a pair of PPT algorithms ($\mathcal{G}, \mathsf{Com}$). The setup algorithm $ck \leftarrow \mathcal{G}(1^\lambda)$ outputs a commitment key $ck$, given a security parameter $\lambda$. The commitment key defines a message space  $\mathcal{M}_{ck}$, a randomness space  $\mathcal{R}_{ck}$, and a commitment space $\mathcal{C}_{ck}$. The commitment algorithm $\mathsf{Com}$ together with $ck$ defines a function $\mathsf{Com}_{ck}:\mathcal{M}_{ck} \times \mathcal{R}_{ck} \rightarrow \mathcal{C}_{ck}$. Given a message $\mu \in \mathcal{M}_{ck}$ the sender picks random value $r \in \mathcal{R}_{ck}$ and calculates the commitment $c=\mathsf{Com}_{ck}(\mu;r)$.
\end{definition}

\begin{definition}
[Hiding \cite{Groth2015}]
\label{def:hiding}
The commitment scheme ($\mathcal{G}, \mathsf{Com}$) is hiding if a commitment does not reveal any information about the committed value. For all PPT stateful adversaries $\mathcal{A}$, it should hold that
\begin{equation*}
  \Pr \left[
    \begin{aligned}
  ck \leftarrow \mathcal{G}(1^\lambda); (\mu_0, \mu_1) \leftarrow \mathcal{A}(ck);\\ d \leftarrow \{0,1\}; c \leftarrow \mathsf{Com}_{ck}(\mu_d) \colon \\ \mathcal{A}(c) = d
  \end{aligned}
  \right]
  \leq \frac{1}{2} + \omega(\lambda) 
\end{equation*}

where $\mathcal{A}$ chooses $\mu_0, \mu_1 \in \mathcal{M}_{ck}$. Here, $\omega(\lambda)$ denotes a negligible function (usually $\leq1/2^{-\lambda}$) and $\lambda$ is the security parameter. 
If the probability is exactly $\frac{1}{2}$, we say the commitment scheme is perfectly hiding.
\end{definition}

\begin{definition}
[Binding \cite{Groth2015}]
\label{def:Binding}
We say that a commitment scheme ($\mathcal{G}, \mathsf{Com}$) is binding if a commitment can only be opened in one way, i.e., not even the randomness can change, and for all PPT
adversaries $\mathcal{A}$
\begin{equation*}
  \Pr \left[\ 
  \begin{aligned}  
    ck \leftarrow \mathcal{G}(1^\lambda); \\ 
    (\mu_0, r_0, \mu_1,r_1); \leftarrow \mathcal{A}(ck): \\
    (\mu_0,r_0)\neq (\mu_1, r_1) \text{ and }\\
     \mathsf{Com}_{ck}(\mu_0;r_0) = \mathsf{Com}_{ck}(\mu_1;r_1)
    \end{aligned}
  \right] \leq \omega(\lambda)
\end{equation*}

where $\mathcal{A}$ chooses $\mu_0, \mu_1 \in  \mathcal{M}_{ck}$ and $r_0, r_1 \in  \mathcal{R}_{ck}$.

\end{definition}

\begin{definition}
[Additive Homomorphic Commitment \cite{Groth2015}]
\label{AHC}

A commitment scheme ($\mathcal{G}, \mathsf{Com}$) is additive homomorphic if: 
\begin{equation*}
 \mathsf{Com}_{ck}(\mu_0;r_0) \cdot \mathsf{Com}_{ck}(\mu_1;r_1) = \mathsf{Com}_{ck}(\mu_0 + \mu_1;r_0 + r_1)     
\end{equation*}
 for all $\mu_0, \mu_1 \in \mathcal{M}_{ck}$ and $r_0, r_1 \in \mathcal{R}_{ck}$. 
\end{definition}

\subsubsection{Pedersen Commitment}
\label{PC}
The Pedersen commitment scheme~\cite{Pedersen1992} is an example of an additive homomorphic commitment scheme, and we use it to design SePEnTra. The Pedersen commitment scheme is perfectly hiding and computationally binding, assuming the discrete logarithm assumption holds \cite{Groth2015}. 

\section{System Design}
\label{systemdesign}
In this section, first, the used notations are listed. After that, we review the market structure and pricing mechanism of TEM. Finally, a general architecture of SePEnTra with the trading mechanism is discussed in detail.

\subsection{Notations}
\label{notations}
We list all frequently used notations in Table \ref{table1}. 

\begin{table} [ht]
\caption{Notations\label{table1}}
\centering
\resizebox{8.5cm}{!}{
\begin{tabular}{cc}
\toprule
\textbf{Notations} & \textbf{Definitions}\\
\midrule
$N$ & Total number of Transactive Agents (TAs)\\

$n$ & Index of TAs, $n=1,\dots,N$ \\

$T $ & Total number of time slots \\

$\varsigma$ & Maximum number of iterations \\

$k$ & Index of iteration, $k=1,\dots,\chi$  \\

$\epsilon$ & Small positive number to terminate iteration\\

$\gamma$ & Price signal  \\

$E_{n,tot}$ & Energy demand or supply forecast of $TA_n$ \\

$E_{n}$ & $TA_n$'s traded energy (demand/supply) in the P2P market\\

$E_{nN}$ & Share of $E_n$ distributed to $TA_N$  \\

$C_{TA_n}$ & Commitment of $E_n$ \\

$r_{n} : r_{n} \in \mathbb{Z}_p $ & Random number to generate $C_{TA_n}$  \\

$r_{nN}$ & Share of $r_n$ distributed to $TA_N$  \\

$e_{n}$ & Real energy demand or supply of $TA_n$  \\

$e_{nN}$ & Share of $e_n$ distributed to $TA_N$   \\

$\beta$ & Threshold to deviate from demand/supply forecast for $N$ TAs  \\

$\sigma$ & Threshold to deviate from demand/supply forecast for one TA  \\

$\underline{v}_{n}, \overline{v}_{n}$ & Dual variables of min and max boundaries of $TA_n$ in the P2P market  \\

$\psi_n$ and $\chi_n$ & $TA_n$'s preference parameters in the P2P market  \\

$\zeta$ & Positive tuning parameter\\

$\lambda$ & Security parameter\\
\bottomrule
\end{tabular}}
\vspace{-0.6cm}
\end{table}

\subsection{Transactive Energy Market (TEM)}
This section provides a brief introduction to the market structure and pricing mechanisms of TEM.

\subsubsection{Market structure and pricing mechanisms}
TEM is a novel market platform that integrates DERs across the electric grid \cite{Cao2022}. In TEM, prosumers and consumers act as transactive players who can interact and negotiate with each other for trading energy and providing other services following market rules. The goal of the TEM is to balance the dynamic demand and supply of energy while all participants are getting fair incentives. TEM operates two types of markets, the day-ahead market and the intra-day or real-time market. Transactive players trade energy a day or hour before the actual energy generation and delivery in the day-ahead market to minimize future uncertainties and prepare the generators for their future operation. Mainly, sellers and buyers commit a certain amount of energy demand and supply based on their generation and load forecast for a specific period that needs to be fulfilled during actual operation/time. In contrast, intra-day market tries to correct imbalances between demand and supply from the day-ahead market. Hence, the day-ahead market plays a pivotal role in the electricity system as false or wrong energy demand/supply forecast has a significant impact on market authorities and participants, including load shedding or wastage of energy~\cite{DRumpa2021}.

TEM has three types of market structure; (i) full peer-to-peer (P2P) market (ii) community-based P2P market, and (iii) hybrid P2P market. In a Full P2P market \cite{Morstyn2019}, peers can communicate directly without any mediator. A community-based P2P market \cite{Akter2016} has a transactive market operator (TO) who leads and synchronizes a community's trading activities. Market participants interact with each other through TO inside a community and the remainder of the system for trading energy. Hybrid P2P market \cite{SOUSA2019} combines the full and community-based P2P market. TO runs a community-based P2P market inside a community in the lower layer. In the upper layer, communities interact among themselves in a full P2P manner.

The pricing mechanisms of TEM can be broadly classified into auction-based and distributed pricing \cite{Rumpa2021}. In auction-based pricing, all market participants submit their bids, consisting of energy supply or demand and energy price. The market authority/TO determines the clearing price based on total demand and supply where both uniform price and pay-as-bid technique can be considered.  Distributed pricing mechanisms work iteratively. At each iteration, the TO estimates the energy price based on previous data and broadcasts the price signal to all market participants. Each participant calculates the demand/supply based on price and sends it to the TO. The TO updates the price signal based on total demand/supply and transmits it to participants, and the process is continued until convergence is reached. This work considers a community-based P2P market structure with distributed pricing mechanisms for trading energy in TEM. However, our proposed model can be easily adapted to the hybrid P2P market.

\subsection{System Model Overview}
\label{system}

\begin{figure}[!t]
\centering
\includegraphics[width=2in]{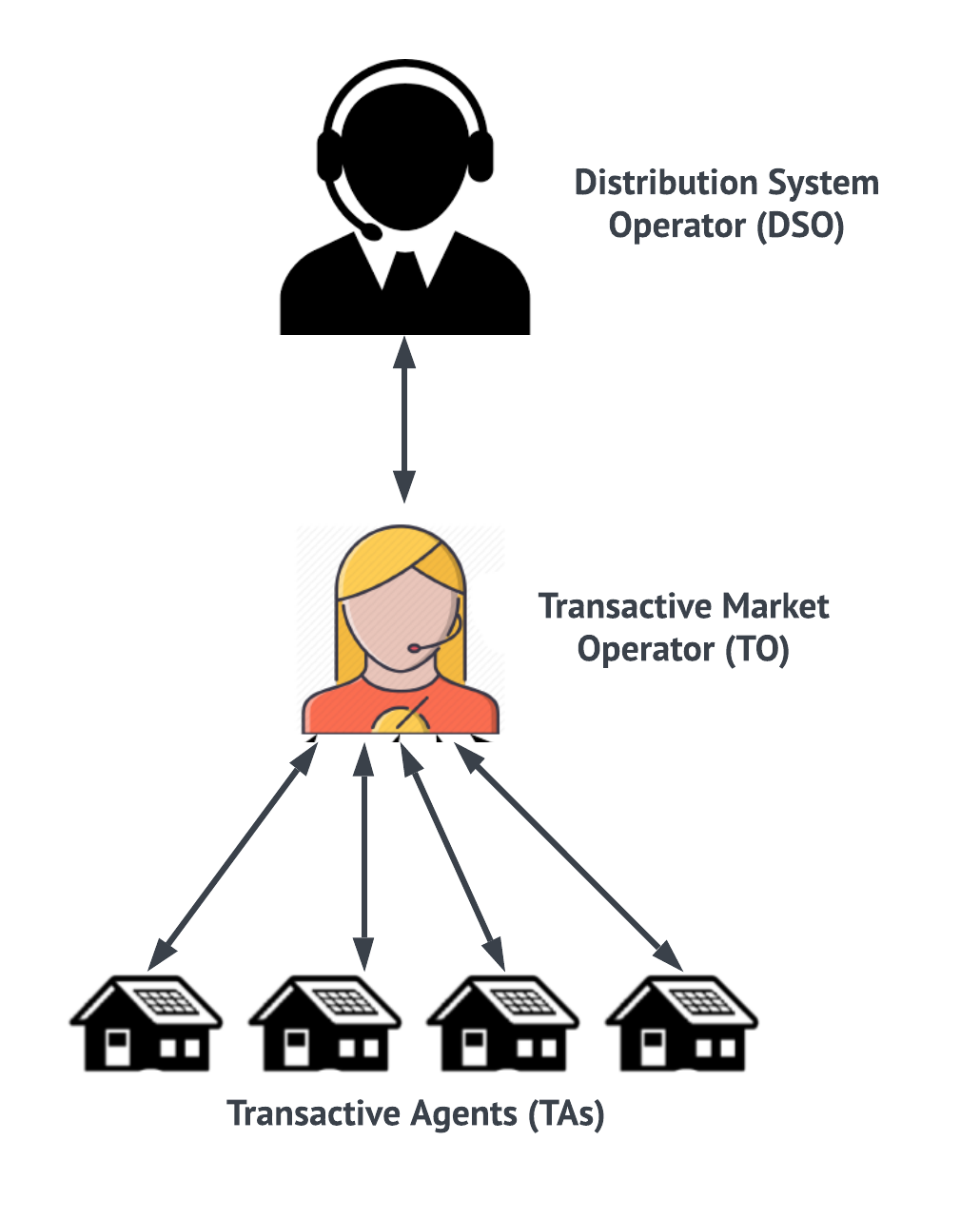}
\caption{System model overview.}
\label{fig1}
\vspace{-0.6cm}
\end{figure}

Our proposed system has three main entities, as shown in Fig. \ref{fig1}. A grid operator or distributed system operator (DSO) monitors the trading activities of a community  and maintains connectivity between the grid and the community. The community consists of N Transactive Agents (TAs). Each TA is equipped with a tamper-proofed smart meter, a HEMS, a set of appliances, and a set of DERs such as solar panels and photovoltaic (PV) batteries. All DERs and appliances are connected with HEMS. HEMS receives energy generation information from connected DERs and determine the optimal starting time of each connected device. HEMS estimates surplus/deficit energy for a particular period using device schedules and DER energy generation forecasts. TA receives information regarding surplus or deficit energy from HEMS. Based on the energy generation or consumption, TA can act as a prosumer/seller or consumer/buyer in TEM. TAs can talk and share information with each other. To better understand our system model, we can compare our community with a university campus or suburb where each building or household represents a TA.

TO coordinates and monitors trading activities inside the community and implements different network constraints to maintain the demand/supply balance. TO is equipped with high computational power hardware to perform necessary computation and has enough storage space. During energy trading among TAs, they need to communicate with others through TO as TO leads and are responsible for market clearing. TO participates in the external market or maintains communication with DSO on behalf of TAs to trade energy from external markets or the main grid. It is noteworthy that TAs can trade any deficit/surplus energy from the grid or external market through TO. However, energy prices at the grid and external market are fixed and can not be negotiated, whereas TAs can negotiate with TO for the price in P2P market. The buying and selling energy price in the P2P is lower and higher, respectively than in the grid and external market. Hence, all TAs try to trade the maximum energy in the P2P market to minimize/maximize their cost/profit.

\textbf{Market Phases:} This work divides the whole market process into two phases. One is the forecast phase, second is the online phase. We divide the whole day or 24 hours into equal time slots such as $t = 1,2,\dots, T$, where $T$ denotes the total number of time slots. The forecast phase occurs before the time addressed by $t$, and the online phase occurs after time $t$. More specifically, if we divide the whole day into 24 one-hour time slots, then 12 AM to 1 AM is addressed by $t=1$. In our scheme, TO runs the forecast phase before 12 AM, where TAs join and make commitments with TO for energy generation or consumption at 12 to 1 AM based on HEMs’ forecast. After 1 AM, the online phase checks the actual energy generation/consumption based on energy data received from the smart meter of each TA.  We present our trading activities and proposed algorithms only for the one-time slot. We ignore time notation for the sake of simplicity. 

\textbf{Overview of Trading Process:} Let $E_{n,tot}$ denotes the energy demand or supply forecast of $TA_n$, for $n \in [1, N]$. We have that $E_{n,tot}>0$, if $TA_n$ has surplus energy, and $E_{n,tot}<0$, if $TA_n$ has deficit energy at a particular period. $E_n$ is defined as $TA_n$'s traded energy in the P2P market. 
Since P2P market prices are beneficial for TAs, each TA aims to trade maximum energy in the P2P market. To clear the P2P market, we consider the trading mechanism introduced in paper \cite{Khorasany2021}. Authors defined a distributed iterative approach using Lagrangian multipliers, and the updating rules for multipliers can be expressed as follows \cite{Mohsen2020}:
\begin{equation}
\label{eq1}
\underline{v}_{n}^{k+1} = [\underline{v}_{n}^{k} - \zeta(E_n^k)]^{+}
\end{equation}
\begin{equation}
\label{eq2}
\overline{v}_{n}^{k+1} = [\overline{v}_{n}^{k} + \zeta(E_n^k - E_{n,tot})]^{+}
\end{equation}
\begin{equation}
\label{eq3}
E_{n}^{k+1} = [E_{n}^{k} + \zeta(\frac{\psi_n-\gamma_{k}+\underline{v}_{n}^{k+1}-\overline{v}_{n}^{k+1}}{\chi_n}-E_{n}^{k})]^{+}
\end{equation}
\begin{equation}
\label{eq4}
\gamma_{k+1} = [\gamma_{k} + \zeta(\sum _{n=1}^{N}E_n^{k+1})]^{+}
\end{equation}
\begin{equation}
\label{eq5}
|\gamma_{k+1} - \gamma_{k} |< 0
\end{equation}
\begin{equation}
\label{eq6}
k>\varsigma,
\end{equation}
where $\underline{v}_{n}$ and $\overline{v}_{n}$ are dual variables of minimum and maximum boundaries of $TA_n$ in the P2P market, $\zeta$ is a small and positive tuning parameter, $\psi_n$ and $\chi_n$ are two preference parameters of $TA_n$, and $[.]^+$ represents $\max(0,.)$. Initially, TO starts the P2P market by sending an initial price signal $\gamma_k$ based on the market forecast to all TAs. Here, $k$ is the iteration number, and $k \geq 1$. Each TA updates $\underline{v}_{n}^{k+1}$ and $\overline{v}_{n}^{k+1}$ using \eqref{eq1} and \eqref{eq2} and use them to update $E_{n}^{k+1}$ following \eqref{eq3}. The updated $E_{n}^{k+1}$ is sent to TO. After receiving $E_{n}^{k+1}$ from $N$ TAs, TO updates its price signal $\gamma_{k+1}$  using \eqref{eq4} and sent the new price signal $\gamma_{k+1}$ to all TAs. This iterative process is stopped when 2 convergence criteria $\gamma_{k+1} - \gamma_k < \epsilon$ and $k>\varsigma$ are met. Here, $\epsilon$ = small positive number to indicate the algorithm termination and $\varsigma$= maximum number of iterations. The price at convergence represents the clearing price or P2P market price for all TAs. 

\section{Threat Model and Security Goals}
\label{threat model}

This section discusses the security and privacy issues associated with our system under the threat model. Then, we introduce the security goals of our proposed scheme.

\subsection{Threat Model}
\label{tm}

The effectiveness of the energy trading process of TEM heavily depends on accurate and timely data originating and propagating through different system components. It helps the system to generate an optimal solution for all market participants. However, distributed and heterogeneous nature of TEM allows external and internal users to execute a number of cyberattacks targeting data accuracy through system components. This paper considers a number of attack scenarios conducted by adversaries through different system components such as communication channels, HEMs, smart appliances, and IoT-integrated DERs.

Our proposed model assumes that the DSO is a trusted entity, whereas the TO is semi-honest. That means TO is an honest but curious adversary. TO correctly follows the trading rules and has no intention of maliciously behaving to produce the wrong result. However, TO may try to gain more information regarding the supply or demand forecast of TAs. TAs (prosumers or consumers) can be malicious. They can inject false energy demand or supply prediction in the forecast phase. Besides malicious TAs, third-party attackers (external attackers) can sniff the communications and alter forecasting data during transmission from one entity to another entity. More specifically, attackers can execute cyberattacks in two ways. First, they can target, hack and take control of different system components to provide false information or modify original data during energy trading. Second, an individual or a group of TAs can intentionally input incorrect or wrong data to receive financial benefits over other TAs. For example, adversaries can launch False Data Injection Attacks (FDIA) by using their own HEMs, smart appliances or manipulating other HEMs to forecast a false demand/supply. Attackers can manipulate communication links between TA and TO during negotiation process to inject inaccurate demand/supply data or replace original data. As TO determines market clearing prices or trade energy from external markets or the grid based on energy prediction of TAs, false or manipulated data could bring significant economic disadvantages for victim TAs while escalating benefits for malicious TAs. Such malicious activities also severely disrupt the regular market operation, leading to market chaos. Hence, generating and sharing correct energy predictions of individual TAs is crucial for market operation and TO. Besides, fair market operation, TO can utilize forecast data to identify malicious TAs or victim TAs by comparing forecast and actual consumption or generation once the exact time is over. It is worth noting that individual TA's energy generation/consumption forecast may not totally match with the actual generation/consumption due to several unavoidable reasons, such as weather fluctuation, the intermittent nature of DERs, user preferences, and emergencies. However, the difference should be kept within a threshold for smooth market operation and to avoid severe consequences.   

While sharing demand/supply forecasts is vital for the above reasons, it is alarming in the context of privacy issues. Knowledge of energy prediction data can reveal TA’s sensitive information to TO or others regarding TA’s presence at home, usage patterns of electrical appliances, and household members in a particular period, which is against the individual’s privacy. Adversaries can eavesdrop during data transmission from TA to TO, or manipulate TO or TO’s database to collect prediction data that they might use to plan for robbery or burglary while TA is out of the home. 

\subsection{Security and Privacy Goals}
\label{securitygoals}

In this work, we set five security goals considering security assumptions and the threat model discussed in \ref{tm}. The goals are listed as follows:

\begin{enumerate}

\item Goal 1 (Privacy against TAs): Not to let TAs learn anything about other TAs’ energy demand/supply prediction or actual consumption/generation information. 
\item Goal 2 (Privacy against TO): To hide each TA's demand or supply forecast from the TO during forecast phase. 

\item Goal 3 (Confidentiality):  To keep data secure (not disclosed or view original information) from third-party attackers or unauthorized users/TAs during transmission from one entity (TA) to another entity (TO).
\item Goal 4 (Integrity): To prevent manipulation of TAs’ promised energy demand or supply value by third-party attackers, malicious TAs, and TA him/herself. 
\item Goal 5 (Detectability): To detect malicious users whose actual energy generation/consumption deviates from energy forecast beyond a threshold.
\end{enumerate}

\section{SePEnTra: Secure and Privacy-Preserving Energy Trading Mechanism}
\label{sepentra}
\begin{figure}[!ht]
\centering
\includegraphics[width=3.5in]{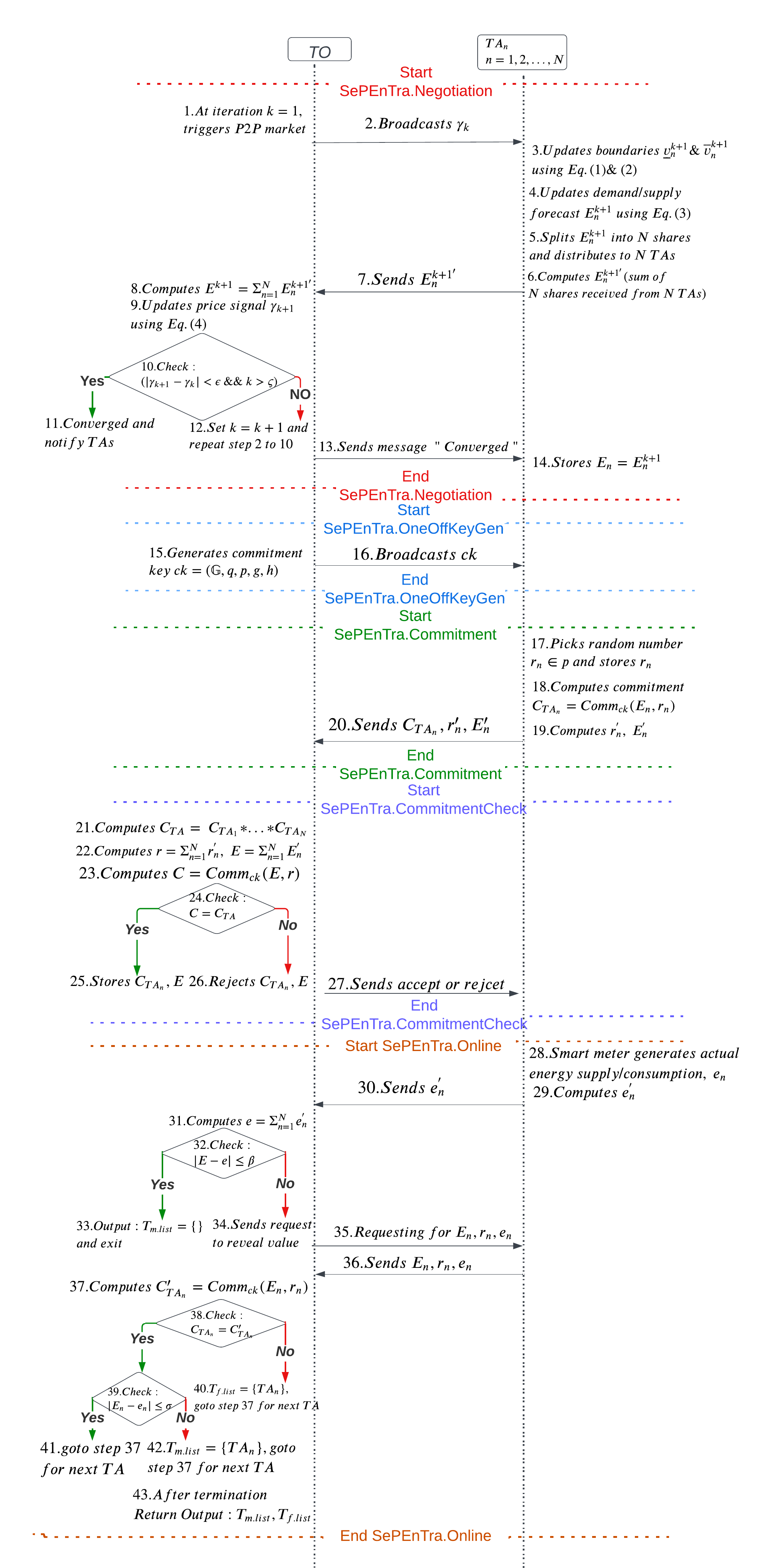}
\caption{SePEnTra Sequence Diagram.}
\label{fig2}
\vspace{-0.6cm}
\end{figure}

This section discusses a detailed framework of our proposed construction, called SePEnTra, with the designed algorithms. SePEnTra consists of five phases: $\Pi_{S}$ = (SePEnTra.Negotiation, SePEnTra.OneOffKeyGen, SePEnTra.Commitment, SePEnTra.CommitmentCheck, SePEnTra.Online). The sequence diagram of our proposed model is depicted in Fig.~\ref{fig2}, which illustrates the protocol sequence. It is important to note that the first four phases of our algorithm are designed for the forecast phase of trading energy, and the fifth phase is developed for the online phase. We refer readers to section \ref{system} for a detailed discussion regarding the forecast and online phase. In this work, SePEnTra.Negotitation, SePEnTra.Commitment, SePEnTra.CommitmentCheck, SePEnTra.Online run for each time slot of a day, whereas SePEnTra.OneOffKeyGen runs once a day. More specifically, if we divide the 24-hour day into 24 1-hour time slots, SePEnTra.Negotitation, SePEnTra.Commitment, SePEnTra.CommitmentCheck, SePEnTra.Online execute 24 times, once per time slot. On the other hand, SePEnTra.OneOffKeyGen perform only once until the key is compromised. In case the key is leaked or hacked by an attacker, SePEnTra.OneOffKeyGen will be executed again to generate a new key. If the key remains uncompromised, TAs reuse the same key for each time interval. That is why we call the phase “One-Off” key generation. 

\textbf{SePEnTra.Negotiation:} TO and TAs perform their respective tasks under this algorithm as depicted in the Algorithm \ref{Negotiation}. This algorithm takes the total number of TAs ($N$), energy demand or supply forecast of individual TA ($E_{n,tot}$), and minimum and maximum boundaries of particular TA to trade energy from the P2P market ($\underline{v}_{n}, \overline{v}_{n}$), two preference parameters of each TA to trade energy from the P2P market ($\psi_n, \chi_n$), small tuning parameter ($\zeta$), initial energy price ($\gamma$), maximum number of iteration ($\varsigma$), and small positive number to terminate iteration ($\epsilon$) as inputs. This algorithm generates the traded energy forecast of individual TA from the P2P market ($E_n$) after a round of negotiations. More specifically, TO starts the first round of the negotiation ($k=1$) of the P2P market by sending $\gamma_k =\gamma$ to TAs. Each $TA_n$, $n\in N$, updates $\underline{v}_{n}^{k+1}, \overline{v}_{n}^{k+1}$ based on input value and using \eqref{eq1} and \eqref{eq2}. Then, $TA_n$ updates $E_n^{k+1}$ using $\underline{v}_{n}^{k+1}, \overline{v}_{n}^{k+1}$ and \eqref{eq3}. $TA_n$ splits $E_n^{k+1}$ to $N$ parts following additive secret sharing scheme described in \ref{ASS}. $TA_n$ keeps the first share of $E_n^{k+1}$ to him/herself and distributes the rest of the share to $(N-1)$ TAs. Each TA holds one share of  $E_n^{k+1}$. The whole process is applicable to all TAs.  Now, each TA has $N$ shares from $N$ TAs. TA sums up $N$ shares received from $N$ users and sends the value to TO. TO adds all value received from $N$ TAs, which is equivalent to $\sum _{n=1}^{N}E_n^{k+1}$ of \eqref{eq4}. TO updates the price signal using \eqref{eq4} and broadcasts it to all TAs. Then, TO runs the next round of negotiation following the same process. This iteration/negotiation is stopped when \eqref{eq5} or \eqref{eq6} are satisfied. After the convergence, TAs receive a notification from TO. The traded energy at convergence (demand/supply) is considered as the final energy demand/supply forecast that TA must fulfil with tolerable deviation during the actual period. $TA_n$ stores $E_n$ at the end of this algorithm. 
\begin{algorithm}[hbt!]
\vspace*{-.05cm}
\footnotesize
\caption{SePEnTra.Negotiation}
\label{Negotiation}
\begin{algorithmic}[1]
\Input{} $N, E_{n,tot},\gamma, \underline{v}_{n}, \overline{v}_{n}, \zeta, \psi_n,  \chi_n,\epsilon, \varsigma$
\Output{}$E_{n}$
\State Set iteration number $k\gets1$, $\gamma_{k+1}\gets$ a big number,$\gamma_{k}\gets \gamma$ $\underline{v}_{n}^{k}\gets\underline{v}_{n}$, $\overline{v}_{n}^{k}\gets\overline{v}_{n}$ \While{$(|\gamma_{k+1} - \gamma_k| \geq \epsilon$ \&\& $k\leq \varsigma)$}
\State.............TAs perform..............
 \For {$ n = 1,\ldots, N $} 
    \State Update $\underline{v}_{n}^{k+1}$ and $\overline{v}_{n}^{k+1}$ using \eqref{eq1} and \eqref{eq2}.
    \State Compute $E_n^{k+1}$ using \eqref{eq3} 
    \State Split $E_n^{k+1}$ to $E_{n1}^{k+1}$,\ldots,$E_{nN}^{k+1}$ where $E_n^{k+1}\gets \sum_{m=1}^N E_{nm}^{k+1}$ 
     \For {$ i = 1,\ldots, N$}
        \State Distribute $E_{ni}^{k+1}$ to $TA_i$
    \EndFor
\EndFor    
\For {$ n = 1,\ldots, N $} 
\State Compute $E_n^{(k+1)^\prime}\gets \sum_{m=1}^N E_{mn}^{k+1}$ 
\State $TA_n$ sends $E_n^{(k+1)^\prime}$ to TO
\EndFor 
\State...............TO perform...............
\State $E^{k+1}\gets \sum_{m=1}^N E_{m}^{k+1^\prime}$ 
\State Update $\gamma_{k+1}$ using \eqref{eq4} and sends $\gamma_{k+1}$ to TAs
\State Update $k\gets k+1$
\EndWhile
\State $TA_n$ stores $E_{n}\gets E_n^{k+1}$
\end{algorithmic}
\end{algorithm}


\textbf{SePEnTra.OneOffKeyGen:} TO generates commitment key $ck = (\mathbb{G},q,p,g,h)$ in this algorithm by taking security parameter $\lambda$ as input. The steps of this algorithm are depicted in Algorithm~\ref{KeyGen}. TO picks two prime numbers $q, p$ in such a way that $p | q – 1$, which is equivalent to $q = b \cdot p + 1$ for random $b$. Then it computes elements of $\mathbb{G}$ using $\{i^b \mod q | i \in \mathbb{Z}_q\setminus \{0\}\}$ and $\mathbb{G}$ order must be equal to $p$. Since the order of $\mathbb{G}$ is prime, any element of $\mathbb{G}$ except 1 is a generator. TO chooses two numbers from $\mathbb{G}$ as generators $g,h$. TO outputs and broadcasts commitment key $ck = (\mathbb{G},q,p,g,h)$ to all TAs. TO also stores $ck$ for future reference.


\begin{algorithm}[hbt!]
\vspace*{-.05cm}
\small
\caption{SePEnTra.OneOffKeyGen}
\label{KeyGen}

\begin{algorithmic}[1]
\Input{}$1^ \lambda$
\Output{}  $ck = (\mathbb{G},q,p,g,h)$
\State ..............TO perform...............
\State Pick two prime numbers $p, q$ primes such that $q = b \cdot p + 1$ for some random $b$
\For {$ i = 1,\ldots, p $}
        \State $\mathbb{G}= \{ {i^b \mod q | i \in \mathbb{Z}_q\setminus 0} \}$
    \EndFor
\State Select two numbers except 1 from $\mathbb{G}$ as generators $g,h$   
\State Store $ck = (\mathbb{G},q,p,g,h)$ 
\State Broadcast commitment key $ck = (\mathbb{G},q,p,g,h)$ to $N$ TAs
\end{algorithmic}
\end{algorithm}


\textbf{SePEnTra.Commitment:} This algorithm (illustrated in Algorithm \ref{Commitment}) managed by TAs, inputs commitment key $ck = (\mathbb{G},q,p,g,h)$ and outputs $C_{TA_n}, E_{n}^\prime, r_{n}^\prime$. $TA_n$ picks a random number $r_n$ from $\mathbb{Z}_p$ and stores $r_n$. Then it generates the commitment of $E_n \in \mathbb{Z}_p$, using $C_{TA_n} = g^{E_{n}}h^{r_n} \mod q$. It is worth mentioning that, generally, $E_n$ is a real number, not an integer number. Hence, we convert $E_n$ from real to an integer by multiplying a large integer number (in our case $10,000$) and rounding up the number using the ceiling and flooring function. After that, $TA_n$ splits $E_n$ and $r_n$ following an additive secret sharing scheme, distributes one share of them to $N$ TAs, and receives shares from other TAs, as we discussed in the SePEntra.Negotiation. Each TA calculates $\sum_{m=1}^N E_{mn}$ and $\sum_{m=1}^N r_{mn}$ to generate $E_{n}^\prime, r_{n}^\prime$. At the end, $TA_n$ transmits $C_{TA_n}, E_{n}^\prime, r_{n}^\prime$ to TO.

\begin{algorithm}[hbt!]
\vspace*{-.05cm}
\footnotesize
\caption{SePEnTra.Commitment}
\label{Commitment}

\begin{algorithmic}[1]
\Input{}$ck = (\mathbb{G},q,p,g,h)$
\Output{} $C_{TA_n}, E_{n}^\prime, r_n^{\prime}$
\State .............. TAs perform.............
\For{$ n = 1,\ldots, N $}
\State Pick a random number $r_n \in \mathbb{Z}_p$
\State Compute $C_{TA_n} \gets g^{E_{n}}h^{r_n} \mod q$
\State Split $E_{n}$ to $E_{n1},\ldots,E_{nN}$ where $E_{n}\gets \sum_{m=1}^N E_{nm}$ 
\State Split $r_{n}$ to $r_{n1},\ldots,r_{nN}$ where $r_{n}\gets \sum_{m=1}^N r_{nm}$ 
     \For {$ i = 1,\ldots, N $}
        \State Distribute $E_{ni}$ to $TA_i$
        \State Distribute $r_{ni}$ to $TA_i$
    \EndFor
\EndFor    
\For {$ i = 1,\ldots, N $}
\State Compute $E_{n}^\prime \gets \sum_{m=1}^N E_{mn}$
\State Compute $r_{n}^\prime \gets \sum_{m=1}^N r_{mn}$ 
\State $TA_n$ stores $r_n$ and sends $C_{TA_n}, E_{n}^\prime, r_{n}^\prime$ to TO
\EndFor
\end{algorithmic}
\end{algorithm}


\textbf{SePEnTra.CommitmentCheck:} TO runs this algorithm by receiving $C_{TA_n},
E_{n}^\prime, r_n^{\prime}$ inputs from $N$ TAs, and notifies TAs about acceptance or rejection of inputs. To begin with, TO computes $C_{TA} = \prod_{m=1}^N C_{TA_m}$. TO also calculates $E = \sum_{m=1}^N {E_{m}^\prime}$ and $r = \sum_{m=1}^N {r_{m}^\prime}$. TO computes commitment of $E$ using $C = g^Eh^{r} \mod q$. Then it compares $C$ and $C_{TA}$. The equality should hold according to Definition \ref{AHC} of additive homomorphic property of the Pedersen commitment scheme as TO performs computation over received commitment. Specifically, the multiplication of $N$ TAs individual forecast commitments provides the same result as the commitment of $N$ TAs total forecast according to the additive homomorphic property of the Pedersen commitment scheme. Therefore, if $C=C_{TA}$, TO stores $E$ and the commitment of $N$ users ($C_{TA_1}$ to $C_{TA_N}$) and broadcasts a notification to all TAs. Otherwise, TO rejects commitments and notifies TAs also. This procedure is illustrated in Algorithm \ref{CommitmentCheck}.


\begin{algorithm}[hbt!]
\vspace*{-.1cm}
\small
\caption{SePEnTra.CommitmentCheck}
\label{CommitmentCheck}
\begin{algorithmic}[1]
\Input{} $C_{TA_1},\ldots,C_{TA_N},E_{1}^\prime,\ldots,E_{N}^\prime, r_1^{\prime},\ldots,r_N^{\prime}$
\Output{} Accept or Reject
\State ..............TO perform...............
\State $C_{TA} \gets \prod_{m=1}^N C_{TA_m}$
\State $E \gets \sum_{m=1}^N {E_{m}^\prime}$ 
\State $r \gets \sum_{m=1}^N {r_{m}^\prime}$ 
\State Compute $C \gets g^Eh^{r} \mod q$
\If {$C = C_{TA} $}
\State TO stores $C_{TA_1},\ldots,C_{TA_N},E$
\State Notify  TAs
\Else
\State Reject inputs and notify TAs
\EndIf
\end{algorithmic}
\end{algorithm}
\textbf{SePEnTra.Online:}
This algorithm receives the actual energy generation/consumption value ($e_n$) from each TA's smart meter, a threshold to deviation for $N$ TAs ($\beta$), and a threshold to deviation for each TA ($\sigma$) as inputs. The algorithm outputs two lists; $T_{m.list}$ and $T_{f.list}$. Here, $T_{m.list}$ contains TAs whose actual energy consumption/generation deviates beyond the threshold from their forecast. On the other hand, $T_{f.list}$ contains TAs whose revealing forecasts under SePEnTra.Online is different than the committed forecast under SePEnTra.CommitmentCheck. The steps of SePEnTra.Online are depicted in Algorithm \ref{Online}. While the actual period is over, $TA_n$ receives $e_n$ from the smart meter. Then, $e_n$ is split and shared with $N$ TAs following an additive secret sharing scheme. $TA_n$ sums up the $N$ shares acquired from $N$ TAs, generates $e_n^\prime$, and sends to TO. 

TO aggregates $e_1^\prime$ to $e_N^\prime$ received from $N$ TAs to determine $e$. TO compares $E$ and $e$ to investigate whether the difference lies within the threshold ($\beta$). If the deviation between the total forecast ($E$) and the actual value ($e$) lies within $\beta$, SePEnTra outputs two null lists; $T_{m.list}$ and $T_{f.list}$. Then, the algorithm is terminated immediately. Otherwise, TO requests to all TAs to reveal respective $E_n$, $r_n$, and $e_n$. Using $E_n$, $r_n$, and $ck$, TO calculates $C_{{TA}_n}^{\prime}$ and compares it with $C_{{TA}_n}$, which the $TA_n$ previously sent and TO stored under Algorithm \ref{CommitmentCheck}. 
If $C_{{TA}_n}^{\prime}=C_{{TA}_n}$, TO checks whether $e_n$ deviates from $E_n$ beyond $\sigma$ or not. If deviation goes beyond $\sigma$, TO puts $TA_n$ to $T_{m.list}$ and repeats the process for the next TA. If $C_{{TA}_n}^{\prime}$ does not match with $C_{{TA}_n}$, TO adds $TA_n$ to $T_{f.list}$ and starts the whole process for next TA. While TO completes checking for all TAs and terminates the algorithm, TO outputs $T_{m.list}$ and $T_{f.list}$.
\begin{algorithm}[hbt!]
\caption{SePEnTra.Online}
\label{Online}
\footnotesize
\begin{algorithmic}[1]
\Input{} ${e_1,\dots,e_N,\beta, \sigma}$
\Output{} $T_{m.list}$, $T_{f.list}$
\State ............TAs perform...........

 \For {$ n = 1,\ldots, N $} 
    \State Receive $e_{n}$ from smart meter
    \State Split $e_{n}$ to $e_{n1}$,\ldots,$e_{nN}$ where $e_{n}\gets \sum_{m=1}^N e_{nm}$ 
    \For {$ i = 1,\ldots, N $}
        \State Distribute $e_{ni}$ to $TA_i$
    \EndFor
 \EndFor  
 \For {$ n = 1,\ldots, N $}
\State Compute $e_n^\prime\gets \sum_{m=1}^N e_{mn}$ 
\State $TA_n$ sends $e_n^\prime$ to TO
\EndFor
\State ...............TO perform.............
\State Set $T_{m.list} \gets \{\}$, $T_{f.list} \gets \{\}$
\State $e\gets \sum_{m=1}^N {e_{m}^\prime}$ 
\If {$|E - e |\le \beta $} 
\State exit
\Else 
\State TO sends message to $N$ TAs for revealing value
\EndIf
\State ............TA perform...........
\For {$ n = 1,\ldots, N $} 
\State $TA_n$ sends $E_{n}$, $r_n$, and $e_n$ to TO
\EndFor
\State ...............TO perform.............
\For {$ n = 1,\ldots, N $}
\State Compute $C_{{TA}_n}^{\prime} \gets g^{E_{n}}h^{r_n} \mod q$
\If {$C_{TA_n} = C_{{TA}_n}^{\prime}  $}
    \If{$|E_n-e_n|\leq \sigma$}
    \State $n=n+1$ and goto line 25 
    \Else 
    \State $T_{m.list} \gets \{TA_n\} $
    \State Notify $TA_n$
    \EndIf
\Else 
\State $T_{f.list} \gets \{TA_n\} $
\State Notify $TA_n$
\EndIf
\EndFor
\State \Return  $T_{m.list}$ and $T_{f.list}$ as output
\end{algorithmic}
\end{algorithm}

\section{Security Analysis}
\label{security}
This section conducts an analysis of the security of SePEnTra. We first define the privacy, hiding, and binding property of our protocol, $\Pi_{S}$, and then present theorems and security proofs of our protocol.
\begin{definition}
[Privacy]
\label{def:privacy}
Let $\left[ N \right]$ denote the set $\{ 1,\ldots, N \}$  where $N$ is the number of TAs and $f(E_1,E_2,\ldots,E_N) = \sum_{i=1}^{N}E_{i}$  denote an $N$-input, single-output function, and let $\Pi_{S}$ be an $N$-TA protocol for computing $f$. We denote sequence of TAs input by $\boldsymbol{\Gamma} = (E_1,\ldots,E_N)$, and the set of $N$-random shares of $E_n$ by $Q_n=\{E_{n1},E_{n2},\ldots,E_{n_N}\}$ where $E_n = \sum_{i=1}^{N}E_{ni}$. Let denote the joint protocol view of TAs in subset $\mathcal{I} \subseteq$ $\left[ N \right]$  by  $\text {VIEW }_{\mathcal{I}}^{\Pi_S}(\boldsymbol{\Gamma}) = (\Phi_1,\Phi_2,\ldots,\Phi_N)_{\mathcal{I}}$, and the protocol output by $\text { OUT }^{\Pi_{S}}(\boldsymbol{\Gamma}) = \Phi_1 +\Phi_2+\ldots+\Phi_N$ where $\Phi_n=\sum_{i=1}^{N}E_{in}$, for $1\leq n\leq N$. We say that $\Pi_{S}$ is a $(N-1)$-private protocol for computing $f$ if there exists a probabilistic polynomial-time algorithm $S$, such that, for every $\mathcal{I} \subset$ $\left[ N \right]$ with $\# \mathcal{I} \leq N-1$ and every $\boldsymbol{\Gamma} \in (\{0,1\}^*)^N$, the random variable
\begin{equation*}
\left\langle S (\mathcal{I}, v_\mathcal{I}, f(\boldsymbol{\Gamma})), f(\boldsymbol{\Gamma}) \right\rangle and \left\langle\text {VIEW }_{\mathcal{I}}^{\Pi_{S}}(\boldsymbol{\Gamma}), \text { OUT }^{\Pi_{S}}(\boldsymbol{\Gamma})\right\rangle
\vspace{-0.1cm}
\end{equation*}
are identically distributed, where $v_\mathcal{I}$ denotes the projection of the $N$-ary sequence $v$ on the coordinates in $\mathcal{I}$ and $\# {\mathcal{I}}$ represents the size of set $\mathcal{I}$. 
\end{definition}

\begin{lemma}
\label{lemma1}
$\Pi_{S}$ is homomorphic and composite if additive secret sharing scheme is homomorphic and composite.
\end{lemma}
\begin{IEEEproof} $\Pi_{S}$ uses additive secret sharing scheme to share $E_n$ with other TAs, where $n\in N$. Since sum of shares are shares of sum$(E_1, \dots, E_N)$, additive secret sharing scheme is homomorphic. The homomorphic sharing scheme is composite \cite{Desmedt93} in the strong sense if it is perfect and the revelation of all shadows of $E=E_1 + \dots + E_N$, where $E_1,\dots E_N \in \Gamma$, does not reveal anything more about $(E_1,\dots, E_N)$ than what $E$ does. Thus if the additive secret sharing scheme is homomorphic and composite, $\Pi_{S}$ is also homomorphic and composite.
\end{IEEEproof}

\begin{definition}
[Hiding]
\label{hiding}
Our protocol $\Pi_{S}$ is hiding if the output  $C_{TA_n},{E_n}^\prime,{r_n}^\prime$ do not reveal the original value and if for all PPT stateful adversaries $\mathcal{A}$, it holds that
\vspace{-0.3cm}
\begin{multline*}
  \Pr\biggl[\{C_{TA_n},{E_n}^\prime,{r_n}^\prime\}_{n \in N}\leftarrow \Pi_{S}(N, E_{n,tot},\gamma, \underline{v}_{n}, \overline{v}_{n}, \\ \zeta, \psi_n,  \chi_n,\epsilon, \varsigma); (E_0, E_1)\leftarrow  \mathcal{A}(C_{TA_n},{E_n}^\prime,{r_n}^\prime); d \leftarrow \{0,1\}; \\ r \leftarrow \mathbb{Z}_p;  c \leftarrow \mathsf{Com}_{ck}(E_d;r) \colon \mathcal{A}(c) = d\biggr] \approx \frac{1}{2}  
  \vspace{-0.4cm}
  \end{multline*}
If the probability is exactly $\frac{1}{2}$, we say that our protocol $\Pi_{S}$ is perfectly hiding.
\end{definition}

\begin{definition}
[Binding]
\label{binding}
Our proposed protocol $\Pi_{S}$ is binding if commitment $C_{TA_n}$ can only be opened in one way i.e., not even the randomness can change and if for all PPT stateful adversaries $\mathcal{A}$, it holds the following approximation
\begin{multline*}
     \Pr\biggl[\{ C_{TA_n},{E_n}^\prime,{r_n}^\prime\}_{n \in N} \leftarrow  \Pi_{S}(N, E_{n,tot},\gamma, \underline{v}_{n}, \overline{v}_{n}, \zeta, \\ \psi_n,  \chi_n,\epsilon, \varsigma);  ( E_0,r_0)\neq (E_1, r_1) ;
    (E_0, r_0, E_1,r_1) \leftarrow  \\ \mathcal{A}(C_{TA_n},{E_n}^\prime,{r_n}^\prime);  \mathsf{Com}_{ck}(E_0;r_0) = \mathsf{Com}_{ck}(E_1;r_1)
 \biggr] \approx 0
 \end{multline*}
\end{definition}

\begin{theorem}
\label{theorem1}
Our protocol $\Pi_{S}$ is $(N-1)$-private in the sense of Definition \ref{def:privacy}, if it is homomorphic and composite, and Pedersen commitment scheme \cite{Pedersen1992} is additive homomorphic. 
\end{theorem}
\begin{IEEEproof} As per Lemma \ref{lemma1}, $\Pi_{S}$ is 
homomorphic and composite.
Since $\Pi_{S}$ uses Pedersen commitment \cite{Pedersen1992} to commit $E_n$ and Pedersen commitment \cite{Pedersen1992} is additive homomorphic according to Definition \ref{AHC}, we can say that our protocol $\Pi_{S}$ is also additive homomorphic.
Thus, if additive secret sharing scheme is homomorphic and composite, and Perdersen commitment scheme is additive homomorphic, our protocol $\Pi_{S}$ is $(N-1)$-private in the sense of Definition \ref{def:privacy}.
\end{IEEEproof}

\begin{theorem}
\label{theorem2}
Our protocol $\Pi_{S}$ is hiding in the sense of Definition \ref{def:hiding}, if Pedersen commitment scheme \cite{Pedersen1992} is hiding and homomorphic sharing scheme \cite{Benaloh1987} is composite.
\end{theorem}
\begin{IEEEproof} We prove the hiding property of our protocol $\Pi_{S}$ through a series of games. Let denote the $i$-th game as $\mathsf{Game}_{i}$.

$\mathsf{Game}_{0}$: Our protocol $\Pi_{S}$ outputs $C_{TA_n}, E_{n}^\prime$, and $r_{n}^\prime$ where $n\in N$ for given inputs. In this game, we replace $E_{n}^\prime$ as a random number. Our scheme uses a homomorphic sharing scheme \cite{Benaloh1987} to share energy demand/supply prediction of $TA_n$ with other TAs. Since the homomorphic sharing scheme is composite \cite{Benaloh1987}, $E_{n}^\prime$ gives no additional information to adversaries $\mathcal{A}$ to break the hiding property of protocol $\Pi_{S}$. \newline
$\mathsf{Game}_{1}$: The only difference of this game from $\mathsf{Game}_0$ is that $r_{n}^\prime$ is replaced by a random number here. Like $\mathsf{Game}_0$, $r_{n}^\prime$ gives no additional information to adversaries $\mathcal{A}$ as homomorphic sharing scheme  is composite \cite{Benaloh1987} in our scheme. Thus, we have 
$|\mathsf{Game}_1-\mathsf{Game}_0|\leq \omega_1(\lambda)$,
where $\omega_1$ is a negligible function in $\lambda$.

$\mathsf{Game}_{2}$: In this game, let us consider $\mathcal{A}$ is a PPT adversary who breaks the hiding property of our protocol, $\Pi_{S}$, to reveal the committed value. Then, we can construct a PPT $\mathcal{B}$ that breaks the hiding property of the Pedersen commitment \cite{Pedersen1992}. The Pedersen commitment \cite{Pedersen1992} is perfectly hiding if the discrete log (DLog) is hard. Hence, we cannot have an adversary $\mathcal{B}$ that breaks the hiding property of the Pedersen commitment \cite{Pedersen1992}. Our protocol $\Pi_{S}$ use Pedersen commitment \cite{Pedersen1992} to commit $TA_n$'s energy demand/supply prediction, $E_n$. Thus, if the Pedersen commitment \cite{Pedersen1992} is hiding, our protocol $\Pi_{S}$ is also hiding. As $\mathsf{Game}_{1}$ is computationally indistinguishable from $\mathsf{Game}_{2}$, we can say that
$|\mathsf{Game}_2-\mathsf{Game}_1|\leq \omega_2(\lambda)$,
where $\omega_2$ is a negligible in $\lambda$. Hence, we get
\begin{equation*}
|\mathsf{Game}_2-\mathsf{Game}_0|\leq \omega_1(\lambda)+\omega_2(\lambda)\leq \omega(\lambda).
\end{equation*}
\end{IEEEproof}
Therefore, if the security of the additive secret sharing scheme and Pedersen commitment scheme holds, the security of our scheme is guaranteed.
\begin{theorem}\label{theorem3}
Our protocol $\Pi_{S}$ is binding as in Definition ~\ref{def:Binding}, if the Pedersen commitment scheme \cite{Pedersen1992} is binding.
\end{theorem}
\begin{IEEEproof} 
Let us assume that $\mathcal{A}$ is a PPT adversary who breaks the binding property of our protocol, $\Pi_{S}$. Then, we can construct a PPT $\mathcal{B}$ that breaks the binding property of the Pedersen commitment \cite{Pedersen1992}. The Pedersen commitment \cite{Pedersen1992} is strongly binding if DLog is hard. Hence, we cannot have an adversary $\mathcal{B}$ that breaks the binding property of the Pedersen commitment. In our protocol $\Pi_{S}$, we use Pedersen commitment to commit $TA_n$'s energy demand/supply prediction, $E_n$. Thus, if the Pedersen commitment \cite{Pedersen1992} is binding, our protocol $\Pi_{S}$ is binding.
\end{IEEEproof}

As per Theorem \ref{theorem1}, our protocol $\Pi_{S}$ is $(N-1)$-private. Except $TA_n$, other TAs or TO not learn anything about $TA_n$’s energy demand/supply forecast ($E_n$) or actual consumption/generation ($e_n$) information. Hence, our privacy goals (Goal 1 and Goal 2) are achieved. In Theorem~\ref{theorem2}, we prove that our protocol $\Pi_{S}$ is hiding. Hence, TA’s energy forecast is not disclosed to third-party attackers or unauthorized users/TAs, and the confidentiality (Goal 3) of our scheme is preserved. Our protocol $\Pi_{S}$ is binding as per Theorem \ref{theorem3}. Adversaries cannot open TA’s commitment with different randomness. Hence, attackers, including third-party and unauthorized users, can not manipulate the energy forecast, and Goal 4 (Integrity) of our scheme is achieved. In SePEnTra. CommitmentCheck, TO stores the commitment of individual TA ($C_{TA_n}$) and the sum of the energy forecast ($E$) of all TAs after verification according to the additive homomorphic property of Pedersen commitment. In the online phase, TO receives the actual energy consumption/supply ($e$) of all TAs. TO compares $E$ and $e$ considering the deviation threshold $\beta$. If the deviation goes beyond the $\beta$, TO requests to reveal the individual’s promised energy supply/consumption value ($E_n$), random number ($r_n$) to compute commitment, and actual energy supply/consumption ($e_n$) at that particular period. Using those values, TO verifies the revealed value by computing commitment on its own and comparing it with previously stored TA’s commitment. While TA passed the verification, TO compares the revealed energy forecast with actual supply/consumption and detects malicious users successfully whose actual value deviates beyond the threshold. Hence, we can say that our proposed scheme achieved Goal 5 listed in Section \ref{securitygoals}.

\section{Analytical Performance Evaluation}
\label{analytical result}
In this section, we investigate and present the following three aspects to analyze the efficacy of SepEnTra.  
\begin{itemize}
\item{Computational Cost: Time needed to execute each phase of SePEnTra}
\item{Communication Overhead: The data transmitted from one TA to other TAs and TO in each phase (for TA).
The data transmitted from TO to TA(s) in each phase (for TO).}
\item{Storage Analysis: Storage space needed for each TA and TO during each individual phase of SePEnTra.}
\end{itemize}
The complexity is calculated for a one-time slot. Notations used to estimate complexity are listed in Table~\ref{table2}. The computational, communication, and storage size complexity of each TA and TO for SePEnTras' phases are shown in Table~\ref{table3}.

\subsection{Computational Cost}
\label{computational}
\subsubsection{SePEnTra.Negotiation} In each round/iteration of SePEnTra.Negotiation, $TA_n$ estimates $\underline{v}_{n}, \overline{v}_{n}$, $E_n$, splits $E_n$ into $N$ shares, and calculates $E_n^\prime$ (summation of $E_{1n},\ldots,E_{Nn}$). Time taken to compute boundaries, energy prediction, split a secret into $N$ shares, and summation of $N$ numbers are $\tau_v$, $\tau_E$ $\tau_S$, and $\tau_A$, respectively. Therefore, the computational cost of SePEnTra.Negotiation for TA is $\tau_v+\tau_E+\tau_S+\tau_A$. For the worst-case scenario, when SePEnTra.Negotiation iterates $\varsigma$ times; the computational complexity of one round multiplied by the maximum number of iterations ($\varsigma$); i.e., $\varsigma \cdot (\tau_v+\tau_E+\tau_S+\tau_A$).

TO adds $N$ numbers to get $E$, calculates the price signal $\lambda$ and checks termination criteria in each round of SePEnTra.Negotiation. Thus, the computational complexity for TO is $\tau_A+\tau_{\gamma}+\tau_t$. For the worst-case scenario, when SePEnTra.Negotiation iterates $\varsigma$ times; the computational cost for TO equals $\varsigma \cdot (\tau_A+\tau_{\gamma}+\tau_t)$.

\subsubsection{SePEnTra.OneOffKeyGen} TO generates cyclic group $\mathbb{G}$ of prime order $p$ in this phase. Hence, computational cost of SePEnTra.OneOffKeyGen is $p\cdot\tau_{\mathbb{G}}$ as time needed to generate one element of $\mathbb{G}$ is $\tau_{\mathbb{G}}$.

\subsubsection{SePEnTra.Commitment} In SePEnTra.Commitment, $TA_n$ commits $E_n$, splits $E_n$ and $r_n$ into $N$ shares, and sums $N$ numbers twice. $TA_n$ takes $\tau_C$, $\tau_S$, $\tau_A$ to commit a value, split a number into $N$ shares and add $N$ numbers, respectively. Thus, TA's computational cost is $\tau_C+2\tau_S+2\tau_A$.

\subsubsection{SePEnTra.CommitmentCheck} During the commitment check, TA has no computation responsibility. On the other hand, TO multiplies $N$ commitments and adds $N$ numbers twice to estimate $C_{TA}, E$, and $r$. Moreover, TO commits a value and compares two commitments in this phase. Hence, the computational complexity for TO is $\tau_M+2\tau_A+\tau_C+\tau_D$.

\subsubsection{SePEnTra.Online} In SePEnTra.Online, $TA_n$ divides $e_n$ into $N$ shares and adds $N$ shares received from $N$ TAs. Therefore, TA's computational cost equals $\tau_S+\tau_A$. 

TO compares total predictions and total consumption/supply and individual’s forecast and individual’s consumption/supply in this phase. TO sums up $N$ shares to get $e$, compares with the value of the total predictions ($E$), estimates the commitment for $N$ TAs after revealing their committed value to verify the commitment of each TA, and compares individual's forecast ($E_n$) and actual supply/consumption ($e_n$). Thus, the computational cost of TO is $\tau_A+N\tau_C+(2N+1)\cdot\tau_D$. 

\vspace{-0.6cm}

\subsection{Communication Overhead} 
\label{communication cost}

\subsubsection{SePEnTra.Negotiation} $TA_n$ transmits $(N-1)$ shares of $E_n$ to $(N-1)$ TAs and $E_n^\prime$ to TO once in a round. We define the size of one share of $E_n$ as $\ell^{\prime}_{E}$ and the size of $E_n^\prime$ is $\ell_{E^\prime}$. Thus, the communication overhead of TA is $(N-1) \cdot \ell^{\prime}_{E}+\ell_{E^\prime}$. For the worst-case scenario, when SePEnTra.Negotiation iterates $\varsigma$ times; the computational complexity of one round multiplied by the maximum number of iterations ($\varsigma$); i.e., $\varsigma\cdot((N-1)\cdot \ell^{\prime}_{E}+\ell_{E^\prime})$.

TO broadcasts price signals to $N$ TAs. Hence, the communication complexity of TO equals $\ell_\gamma$ for one iteration as the size of the price signal is $\ell_\gamma$. For the worst-case scenario, when SePEnTra.Negotiation iterates $\varsigma$ times; the communication cost for TO equals $\varsigma \cdot\ell_\gamma$.

\subsubsection{SePEnTra.OneOffKeyGen} 
SePEnTra.OneOffKeyGen broadcasts $ck$ to all TAs. $ck$ includes $q$,$p$,$g$,and $h$. Hence, the communication cost of TO is $\ell_q+\ell_p+\ell_g+\ell_h$ in this phase.

\subsubsection{SePEnTra.Commitment} During SePEnTra.Commitment, $TA_n$ sends $(N-1)$ shares of $E_n$ and $r_n$ to $(N-1)$ TAs. Furthermore, $TA_n$ transmits $C_{TA_n}, E_{n}^\prime, r_{n}^\prime$ to TO. Therefore, the communication complexity of TA is $(N-1)\cdot\ell^{\prime}_{E}+(N-1)\cdot\ell^{\prime}_{r}+\ell_C+\ell_{E^\prime}+\ell_{r^\prime}$.

\subsubsection{SePEnTra.CommitmentCheck} In the commitment check phase, TA and TO do not need to send data to each other. Thus, no communication overhead for TA and TO in SePEnTra.CommitmentCheck. 

\subsubsection{SePEnTra.Online} In SePEnTra.Online, $TA_n$ sends $(N-1)$ shares of $e_n$ to $(N-1)$ TAs and ${e_n^\prime}$ to TO. Moreover, $TA_n$ reveals the committed value ($E_n$), the random number to commit the value ($r_n$), and actual energy supply/consumption ($e_n$) to TO. We define the size of ${e_n^\prime}$, $E_n$, $r_n$, $e_n$ and one share of $e_n$, as  $\ell_{e^\prime}$, $\ell_E$, $\ell_r$, $\ell_e$ and $\ell^{\prime}_{e}$, respectively. Hence, the communication complexity of TA equals $(N-1) \cdot \ell^{\prime}_{e}+\ell_{e^\prime}+\ell_E+\ell_r+\ell_e$.
As TO does not send data to TA except for requests to reveal the original value, which is negligible, there is no communication overhead for TO.

\subsection{Storage Analysis} 
\label{storage}
In SePEnTra.Negotiation, $TA_n$ stores $E_n$. The size of $E_n$ equals $\ell_E$. Hence, the storage size is $\ell_E$. Note that TA stores $E_n$ after the termination of SePEnTra.Negotiation. Hence, the iteration number does not affect the storage size. 
TO stores $ck$ in SePEnTra.OneOffKeyGen. The size of $ck$ is $\ell_q+\ell_p+\ell_g+\ell_h$, which is the summation of the size of $q,p,g,h$.
In SePEnTra.Commitment, Each TA stores a random number $r$ where $r\in \mathbb{Z_p}$ in this phase. The size of a random number equals $\ell_r$. Thus, the storage size of SePEnTra.Commitment is $\ell_r$.
In SePEnTra.CommitmentCheck, TO's storage size is $N\ell_C + \ell_E$, which is the size of a commitment ($\ell_C$) multiplied by the number of TAs ($N$) plus the size of total energy forecasts ($\ell_E$). 


\begin{table} [ht]
\caption{Notations\label{table2}}
\centering
\resizebox{8.5cm}{!} {
\begin{tabular}{cc}
\toprule
\textbf{Notations} & \textbf{Definitions}\\
\midrule
$\tau_v$ & Time needed to compute energy trading boundaries in P2P market\\

$\tau_E$ & Time needed to compute an energy supply/demand forecast \\

$\tau_S$ & Time needed to split a secret into $N$ shares \\

$\tau_A$ & Time needed to sum $N$ shares of energy supply/demand forecasts \\

$\tau_{\lambda}$ & Time needed to compute a price signal \\

$\tau_t$ & Time needed to check termination criteria\\

$\tau_{\mathbb{G}}$ & Time needed to generate an element of $\mathbb{G}$  \\

$\tau_{C}$ & Time needed to compute a commitment \\

$\tau_{M}$ & Time needed to multiply $N$ commitments \\

$\tau_{D}$ & Time needed to compare two values \\

$\ell_x$ & Size (bit-length) of $x$\\

$\ell^{\prime}_{x}$ & Size (bit-length) of one share of $x$ \\

$\ell_{x^\prime}$ & Size of the sum of $N$ shares of $x$\\
\bottomrule
\end{tabular}}
\vspace{-0.6cm}
\end{table}

\begin{table*}[ht]
\caption{Computational and communication Costs of SePEnTra.}
\footnotesize
    \centering
    \resizebox{\textwidth}{!} {
\begin{tabular}{*{5}{c}}
    \toprule
\textbf{Algorithm Name}    & \textbf{TA/TO}  & \textbf{Computational Cost}  & \textbf{Communication Cost}  & \textbf{Storage Size} \\
    \midrule
SePEnTra.Negotiation (For one iteration) & TA    & $\tau_v+\tau_E+\tau_S+\tau_A$         & $(N-1)\cdot \ell^{\prime}_{E}+\ell_{E^\prime}$  & $\ell_{E}$        \\ \cline{2-5}
        
            & TO    & $\tau_A+\tau_{\gamma}+\tau_t$           &  $\ell_\gamma$  & -           \\

SePEnTra.Negotiation (For $\varsigma$ iterations) & TA    & $\varsigma \cdot (\tau_v+\tau_E+\tau_S+\tau_A)$         & $\varsigma \cdot((N-1) \cdot \ell^{\prime}_{E}+\ell_{E^\prime})$   & $\ell_{E}$       \\ \cline{2-5}
        
            & TO    & $\varsigma \cdot (\tau_A+\tau_{\gamma}+\tau_t)$           &  $\varsigma \cdot \ell_\gamma$    & -        \\

SePEnTra.OneOffKeyGen           & TA    & -       & -    &-      \\ \cline{2-5}
        
            & TO    & $p \cdot \tau_\mathbb{G}$          & $\ell_q+\ell_p+\ell_g+\ell_h$      & $\ell_q+\ell_p+\ell_g+\ell_h$     \\

 SePEnTra.Commitment           & TA    & $\tau_C+2\tau_S+2\tau_A$         & $(N-1) \cdot \ell^{\prime}_{E}+(N-1) \cdot \ell^{\prime}_{r}+\ell_C+\ell_{E^\prime}+\ell_{r^\prime}$   & $\ell_r$        \\ \cline{2-5}
        
            & TO    & -          & -      & -    \\  
   
SePEnTra.CommitmentCheck           & TA   & -         & -    & -     \\ \cline{2-5}
        
            & TO    & $\tau_M+2\tau_A+\tau_C+\tau_D$          & -    & $N\ell_C+\ell_E$      \\  
    
 SePEnTra.Online          & TA   & $\tau_S+\tau_A$         & $(N-1) \cdot \ell^{\prime}_{e}+\ell_{e^\prime}+\ell_E+\ell_r+\ell_e$        & -    \\ \cline{2-5}
        
            & TO    & $\tau_A+N\tau_C+(N+1) \cdot \tau_D$          &   -    & -        \\  
    \bottomrule
    
\end{tabular}
}
\label{table3}
\vspace{-0.6cm}
\end{table*}

\section{Experimental results}
\label{Experimental results}

This section demonstrates the implementation results of SepEnTra with various numbers of TAs and different orders of $q$ and $p$. We develop our source code in Java, and our machine for running it is an Intel(R) Core(TM) i5-8265U CPU @ 1.60GHz  processor with 16 GB RAM running Windows 10. This work evaluates the computational cost, communication overhead, and storage analysis to analyze the efficacy of our proposed method. We compare these three parameters of SePEnTra with the same TEM framework without adopting any security measures. Note that as discussed and showed in Section \ref{analytical result} and Table \ref{table3}, some phases of SePEnTra do not require any data transmission between TA and TA or TA and TO. In some phases, TA or TO does not need to store any data. Hence, the communication cost or storage size for TA or TO of some phases do not reflect in our results.

We implement energy trading of SepEnTra for the one-time slot, which is assumed to be one hour (1h). To set up the market framework, a community with an equal number of prosumers/sellers and consumers/buyers is considered. We follow \cite{Khorasany2021} and \cite{MKhorasany2020} to set all major parameters of each TA and TO. We use Miller–Rabin primality test \cite{RABIN1980} with 5,000 rounds to generate all prime numbers during SePEnTra implementation. Table \ref{table4} summarizes the range of values for different input parameters.

\begin{table} [ht]
\caption{Input Parameters Values \label{table4}}
\centering
\begin{tabular}{cc}
\toprule
\textbf{Parameter} & \textbf{Value}\\
\midrule
$\underline{v}_{n}$ & [0, 5] kWh\\

$\overline{v}_{n}$ & [3, 20] kWh \\

$\chi_n$ & [0.09, 0.1] \textcent/kWh\textsuperscript{2} \\

$\psi_n$  & [24, 38] \textcent/kWh\textsuperscript{2} \\

$E_{n,tot}$ & [0, 20] kW \\

$\gamma$ & 10 \textcent/kWh\\

$\varsigma$ & 100\\

$\epsilon$ & 0.001 \\

$N$ & 100 \\

$O(\log_{2}b)$ & 1000   \\

$O(\log_{2}p)$ & 20   \\

$O(\log_{2}q)$ & 1020  \\
\bottomrule
\end{tabular}
\vspace{-0.6cm}
\end{table}

\subsection{Computational Cost}
\label{Computational Cost}

In this section, we illustrate and analyze the execution time of five phases of SePEnTra and the total execution time of SePEnTra, which excludes the execution time of the SePEnTra.OneOffKeyGen phase with various numbers of TAs, order of $\log_{2}q$ $(O(\log_{2}q))$, and order of $\log_{2}p$ $(O(\log_{2}p))$.

\subsubsection{Effects of the number of TAs}
\label{ComputationalTA}

\begin{figure}[!t]
\centering
\includegraphics[width=\linewidth]{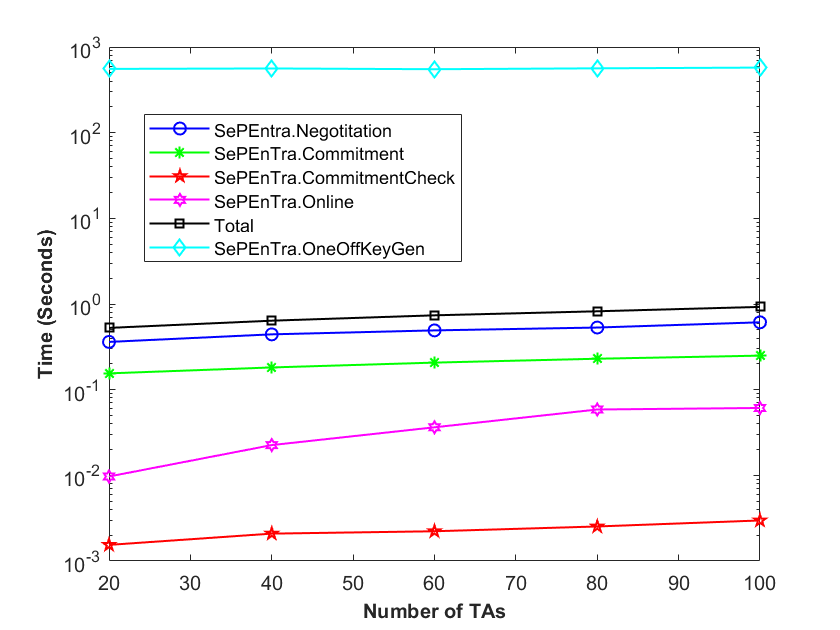}
\caption{Computational time with various number of TAs.}
\label{computational1}
\vspace{-0.6cm}
\end{figure}

We change the number of TAs from 20 to 100 while other input parameters remain the same as in Table \ref{table4}. The execution time of five phases of SePEnTra separately and the total execution time of SePEnTra excludes the SePEnTra.OneOffKeyGen phase is illustrated in Fig. \ref{computational1}. While the number of TAs in the community grows, the computation time of all phases of SePEnTra barely increases. The total execution time of SePEnTra lies under 1 second, even with 100 TAs. The generation of the commitments is computationally very challenging because multiple operations with higher bits of $p, q$ are associated with this. In our proposed construction, SePEnTra.Commitment takes around 250 milliseconds to generate 100 TAs commitment which is time-efficient. Hence, we can say that our proposed model is suitable for large scenarios or communities with a higher number of community members. In our implementation and comparison purposes, we consider the total number of TAs as 100 as default. 

Furthermore, it is evident from Fig.~\ref{computational1} that among the 5 phases of SePEnTra, SePEnTra.OneOffKeyGen has the highest execution time, which is around 560 seconds (9.33 mins). We had $O(\log_{2}q)=1,020$ and $O(\log_{2}p)=20$ to ensure security. As KeyGen executes only once and it is performed by TO, which is equipped with proper hardware and has more computation power, the time is in tolerable range due to providing proper security and privacy of TAs. 

\subsubsection{Effects of $\log_{2}q$}
\label{Computationalq}

\begin{figure}[!t]
\centering
\includegraphics[width=\linewidth]{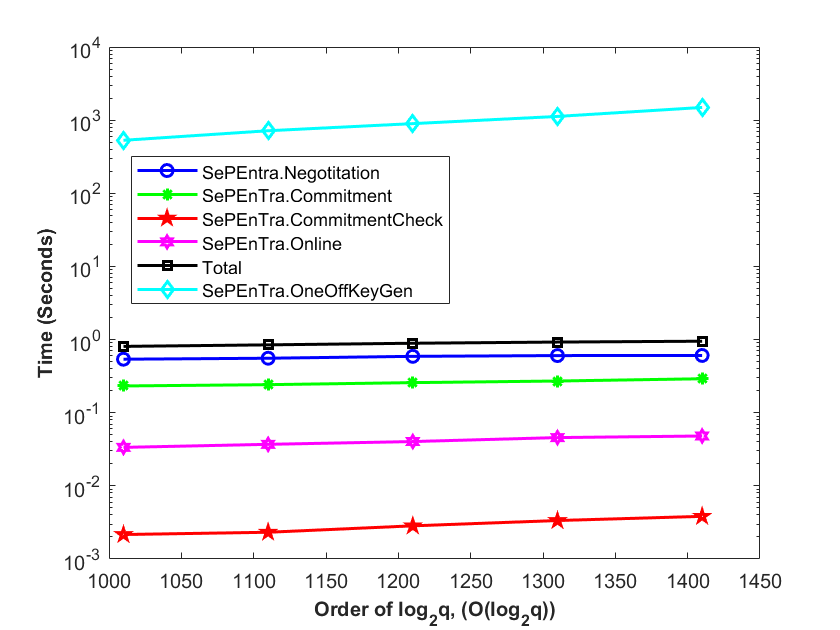}
\caption{Computational time with various order of $\log_{2}q$.}
\label{computational2}
\end{figure}

We consider the order of $\log_{2}b$ from 990 to 1,390 to get a large order of $\log_{2}q$ (1,010 to 1,410) as $q= b \cdot p+1$, and we set $O(\log_{2}p)$ = 20. The computation time of all phases of SePEnTra with varying order of $\log_{2}q$ is depicted in Fig. \ref{computational2}. As seen in Fig. \ref{computational2}, the execution time of SePEnTra rises slightly (a few milliseconds) with the orders of $\log_{2}q$ except for SePEnTra.OneOffKeyGen. In SePEnTra.OneOffKeyGen, the time needed to generate the commitment key by TO is proportional to the order of $\log_{2}q$. TO takes around 529.94 seconds (8.81 mins) to 1,491.711 seconds (24.86 mins) to generate the commitment key while the order of $\log_{2}q$ is 1,010 to 1,410. The time is still suitable for real-time implementation as it is a one-off mechanism and managed by high computational power equipped TO.

\subsubsection{Effects of $\log_{2}p$}
\label{Computationalp}

\begin{figure}[!t]
\centering
\includegraphics[width=\linewidth]{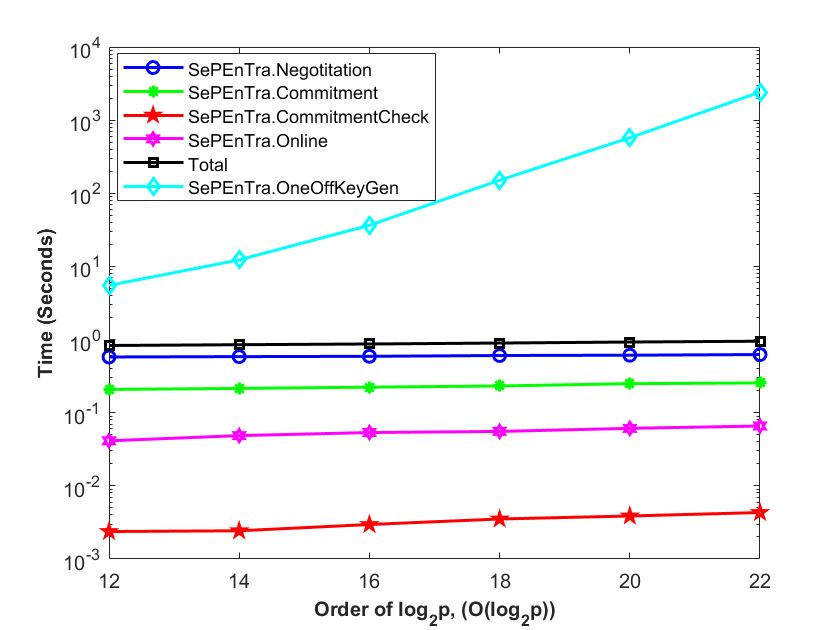}
\caption{Computational time with various order of $\log_{2}p$.}
\label{computational3}
\vspace{-0.6cm}
\end{figure}

In this subsection, the order of $\log_{2}p$ has changed from 12 to 22 to analyze the execution time of SepEnTra. It is evident from Fig. \ref{computational3}, except for SePEnTra.OneOffKeyGen, the execution time of different phases of SePEnTra lies between a few microseconds to milliseconds with various orders of $\log_{2}p$, which is negligible considering the whole large scenario. As SePEnTra.OneOffKeyGen generates a cyclic group $\mathbb{G}$ of prime order $p$, the computational cost of this phase entirely depends on the order of $p$. Therefore, we can see a sharp rise in execution time from 574.8 seconds to 2,416.79 seconds when $O(\log_{2}p)$ grows from 20 to 22 bits. If $O(\log_{2}p)=20$, the cycle group has more than 1 million elements, which is reasonable for security purposes of any large scenario. For $O(\log_{2}p) = 20$ and $O(\log_{2}q) = 1,020$ while $O(\log_{2}b) = 1,000$, the execution time of one-off KeyGen is around 574.876 seconds or 10 mins (see Fig. \ref{computational3}), which is practical due to the sake of security. Hence, during our implementation, we consider $O(\log_{2}p)=20$ as the default. 

\subsection{Communication Overhead}
\label{Communication Overhead}

In this section, we present the communication overhead of different phases of SePEntra, both for TO and TA. We consider the amount of data transferred between TO and TAs as TO’s communication overhead. On the other hand, for TA, the amount of data transmitted between TA and other TAs and TA and TO are considered. 

Table~\ref{table:6} shows the communication overhead when the total number of TAs is 100. The overhead of each TA and TO is less than 1 KB in different phases of SepEnTra except the negotiation phase. In SePEnTra negotiation, this overhead increases if the iteration rounds ($k$) or the total number of TAs ($N$) increases. Here we consider the worst-case scenario, i.e., $k$ = $\varsigma$. Still, the communication overhead is 39.06 KB for each TA, which is negligible in practical scenarios. 

\begin{table}
\caption{Communication cost \& Storage size of each TA and TO while total number of TAs is 100.}
\label{table:6}
\centering
\resizebox{\columnwidth}{!}{%
\begin{tabular}{ccccccc}
\toprule
\multirow{2}{*}{Algorithm }&\multicolumn{2}{c}{Communication Cost (KB)}&\multirow{1}{*}{}&
\multicolumn{2}{c}{Storage Size (KB)}\\
\cline{2-3}\cline{5-6} 
&TA&TO&&TA&TO\\
\midrule
SePEnTra.Negotiation  & 39.06 &0.39 & &0.0039    & -    \\
SePEnTra.OneOffKeyGen & - & 0.376 &&-  & 0.376  \\
SePEnTra.Commitment & 0.905 &-& &0.0039  & - \\
SePEnTra.CommitmentCheck & - &- & &-  & 12.46 \\
SePEnTra.Online & 0.402 &- & &0.0039   &-  \\
\bottomrule
\multicolumn{6}{p{6cm}}{}
\end{tabular}}
\vspace{-0.6cm}
\end{table}

\begin{figure}[!t]
\centering
\includegraphics[width=\linewidth]{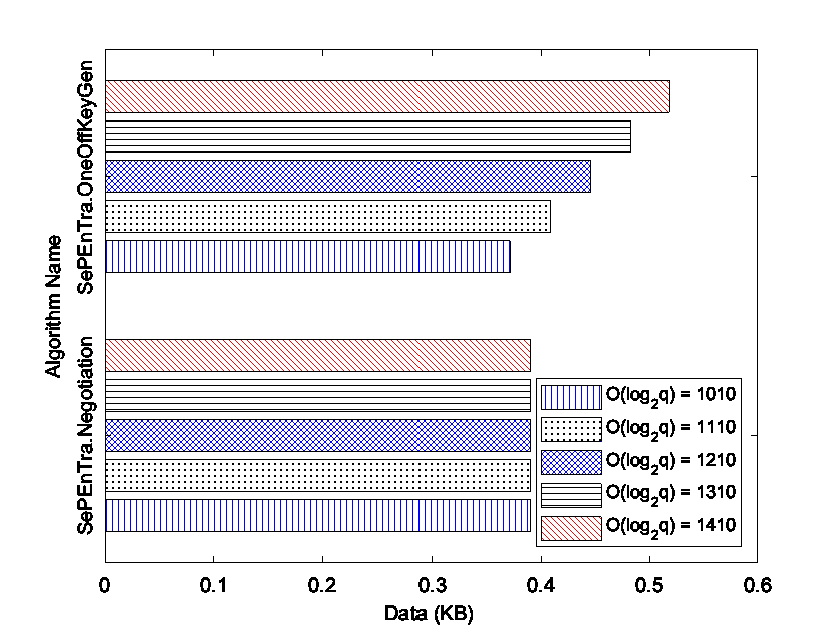} 
\caption{Communication cost of TO with various orders of $\log_{2}q$.}
\label{communication2}
\vspace{-0.6cm}
\end{figure}

Fig. \ref{communication2} represents the communication cost of TO with various orders of $\log_{2}q$. The overhead is constant (0.39 KB) in the SePEnTra.Negotiation as TO only transmits the price signal under this phase. In the one-off KeyGen, this overhead grows linearly with orders of  $\log_{2}q$. The overhead is 0.3723 KB to 0.5188 KB for $\log_{2}q$ as TO needs to broadcast $q$, $p$, $g$, and $h$ to TAs during this phase. We collect communication overhead of TO with various orders of $\log_{2}p$ as well. The result is the same as Fig. \ref{communication2} except for SePEnTra.OneOffKeyGen. During key generation, the overhead changes from 0.3720 KB to 0.3769 KB while the order of $\log_{2}p$ varies from 12 to 22. 

We exclude the communication overhead of TA from Fig. \ref{communication2} as it is constant and the same as Table \ref{table:6}. The reason is that the amount of data transmitted by TA does not relate to orders of $\log_{2}q$ and $\log_{2}p$. Only in SepEnTra.Commitment,  TA sends commitment to TO, which is associated with orders of $\log_{2}q$ and $\log_{2}p$. However, TA's overhead varies from 0.9045 KB to 0.9656 KB (for $O(\log_{2}q)$) and 0.9047 KB to 0.9057 KB (for $O(\log_{2}p)$) in SepEnTra.Commitment, which is insignificant.

\subsection{Storage Analysis}
\label{Storage Analysis}

\begin{figure}[!t]
\centering
\includegraphics[width=\linewidth]{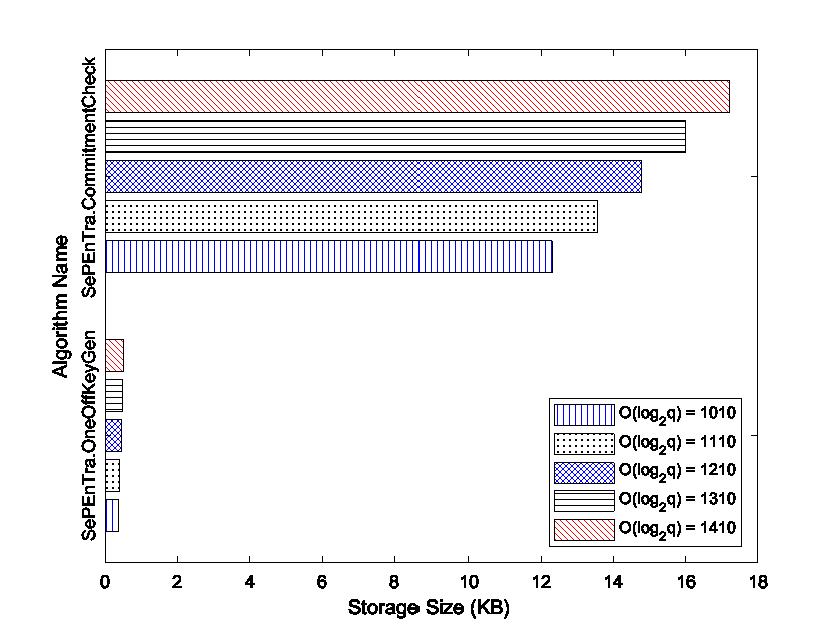} 
\caption{Storage space needed for TO with various orders of $\log_{2}q$.}
\label{storage2}
\vspace{-0.6cm}
\end{figure}

Table \ref{table:6} shows the storage space required for each TA and TO when $N$ =100. In SePEnTra, TA and TO do not require huge storage space. Only in SePEnTra.Commitmentcheck, TO needs to store commitment of $N$ TAs. The length of commitment depends on order $\log_{2}q$. Hence, the storage space increases with higher orders of $\log_{2}q$ as shown in Fig. \ref{storage2}. However, the storage space is less than 18 KB and 1 KB for TO and TA, respectively, in different phases of SePEnTra. Hence, we can say that our proposed model is storage efficient for users and market operators.

\vspace{-0.6cm}

\subsection{Detection Accuracy of SePEnTra}
\label{detection accuracy}

As our fifth goal is detectability where we aim to detect malicious TAs whose actual energy supply/consumption deviates more than the forecast, we measure the accuracy of SePEnTra’s detection mechanism in this section. In SePEnTra.Online, TO performs a two-phase detection mechanism where TO performs the second phase of detection for those TAs who pass the first phase. More specifically, in the first phase, TO detects TAs whose revealed values ($E_n$, $r_n$) under SePEnTra.Online do not match with the committed forecast stored under SePEnTra. CommitmentCheck. First-phase detection aims to verify the revealed value. It is important to note that as TO received the actual energy supply/consumption of each TA generated from the tamper-proofed smart meter, TO excludes $e_n$ from the verification process. After verification, TO detect malicious TAs whose actual supply/consumption ($e_n$) received from TA’s smart meter deviates beyond threshold $\sigma$ than the forecast value ($E_n$) under the second phase. To calculate the accuracy of SePEnTra’s detection mechanisms, we manipulate $e_n$, $E_n$, and $r_n$ of random 15 TAs out of 100. We modify the original values of either $e_n$, $E_n$, or $r_n$ for those TAs by a range of 5\% to 10\%. Then, we run the code 500 times. In every case, TO successfully adds TAs either to $T_{f.list}$ or $T_{m.list}$. If $C_{{TA}_n}^{\prime}$  based on $E_n$ and $r_n$ does not match with previously stored $C_{TA_n}$, TO adds them to $T_{f.list}$. While the difference between $e_n$ and $E_n$ goes beyond $\sigma$, TO adds them to $T_{m.list}$. Hence, we can say that accuracy of the detection mechanism of SePEnTra is 100\%.

\subsection{Performance Comparison}
\label{Performance Comparison}

This subsection compares the performance of SePEnTra with the same market framework without considering security mechanisms. Note that, here without security measures refers to the basic energy trading framework of TEM. More specifically, without considering a secret sharing method and commitment scheme during different phases of energy trading of TEM.  We change the number of TAs from 20 to 100 during computation cost comparison, while for communication overhead and storage analysis, we use 100 TAs with the order of $\log_{2}q$ and $\log_{2}p$ as 1,000 and 20. The rest of the parameters remain unchanged, as shown in Table \ref{table4}. 

\begin{figure}[!t]
\centering
\includegraphics[width=\linewidth]{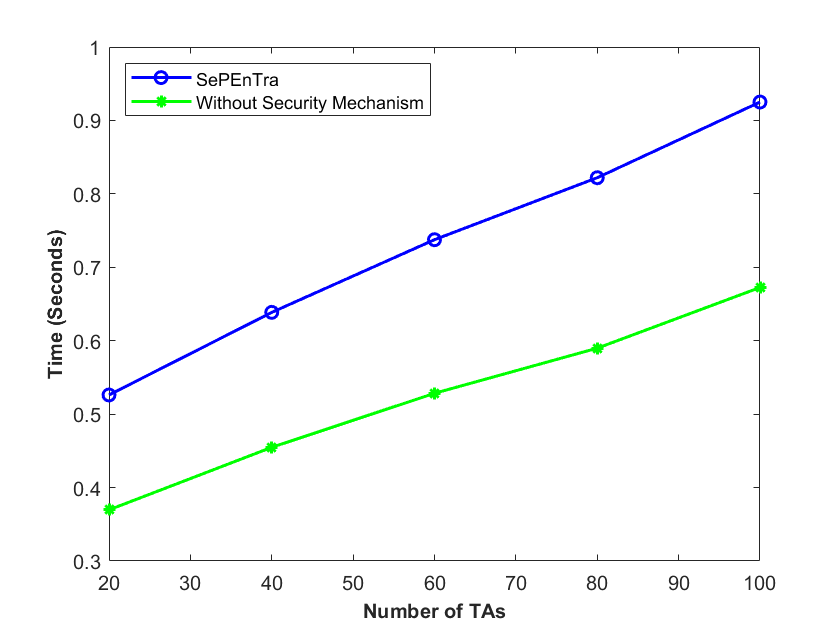} 
\caption{Execution time comparison of SePEnTra (exclude SePEnTra.OneOffKeyGen) with no security measures.}
\label{comparison1}
\vspace{-0.6cm}
\end{figure}

Fig. \ref{comparison1} compares the execution time of SePEnTra with no security measures of TEM. It is worth mentioning that we exclude the execution time of the SePEnTra.OneOffKeyGen phase from SepEnTra for a fair comparison, as this phase is needed to be executed once only. We can see from Fig. \ref{comparison1}, the execution time increases linearly with TAs for both cases. As SePEnTra uses advanced cryptographic primitives in different phases, such as additive secret sharing scheme in negotiation and commitment phases, Pedersen commitment in commitment, commitment check, and online phases, the execution time must be higher than with no security. Fig. \ref{comparison1} also verifies our above statement. Here, the significant point is despite using several cryptographic primitives which involve many computationally demanding operations, the execution time of SePEnTra lies between 0.53 seconds and 0.92 seconds. To ensure security and guarantee privacy, less than 1 second time is trivial in such a large and practical scenario of TEM.

Table \ref{table:7} compares the communication overhead of each TA and TO between SePEnTra and adopting no security techniques. As we stated previously, TA uses additive secret sharing and needs to transmit the share of secrets with other TAs during and after negotiation, which leads to slightly higher communication overhead for each TA in SePEnTra. However, 40 KB data transmission is not a matter for a user of the TEM framework as it provides privacy in addition to security. TO takes only 0.77 KB as it broadcasts $\gamma_k$ and $ck$ in SePEnTra. Without security measures, TA's communication overhead is 0.39 KB as it transmits the whole prediction data to TO.

\begin{table}
\caption{Communication cost of each TA and TO with and without security measures.}
\label{table:7}
\centering
\resizebox{\columnwidth}{!}{%
\begin{tabular}{ccccccc}
\toprule
\multirow{2}{*}{Algorithm }&\multicolumn{2}{c}{Communication Cost (KB)}\\
\cline{2-3} 
&TA&TO\\
\midrule
SePEnTra  & 40.37 &0.77   \\
Without Security Mechanisms & 0.39 & 0.39  \\

\bottomrule
\multicolumn{3}{p{7cm}}{}
\end{tabular}}
\vspace{-0.6cm}
\end{table}

\begin{figure}[!t]
\centering
\includegraphics[width=\linewidth]{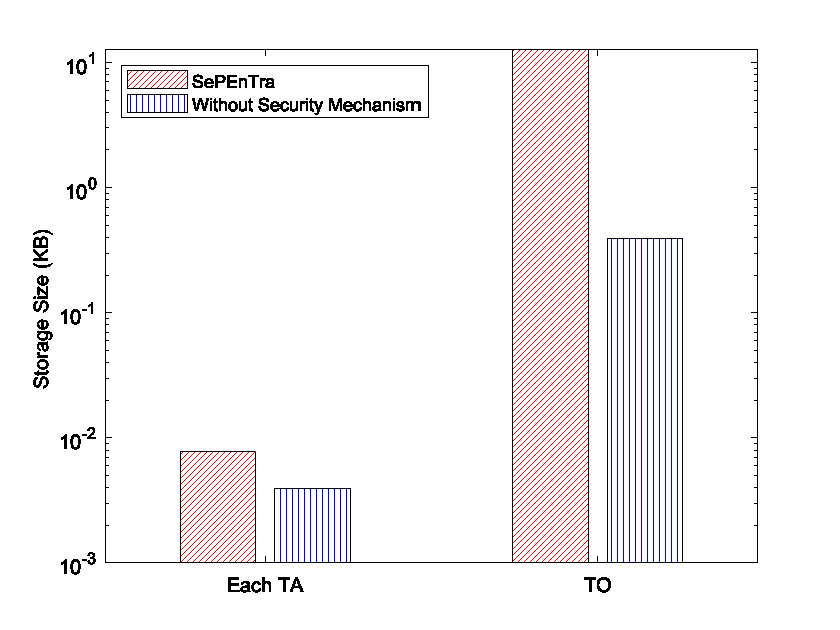} 
\caption{Storage analysis between SePEnTra and without security mechanisms of the TEM}
\label{comparison2}
\vspace{-0.6cm}
\end{figure}

The storage size comparison is shown in Fig.\ref{comparison2}.  TA requires minimal storage space in both cases, which is negligible. Without security mechanisms, TO stores TA’s energy forecast after the commitment phase, and the required space is 0.39 KB (for 100 TAs). On the other hand, SePEnTra provides both security and privacy by storing the commitment of TA’s energy forecast ($C_{TA_n}$). Commitment length depends on the order of $\log_{2}q$. Here, TO requires 12.83 KB of space to store 100 TAs commitments, and the length of each commitment is 1,020 bits. Hene, we can undoubtedly say that SePEnTra is storage efficient as well.

\vspace{-0.4cm}

\section{Conclusion}
\label{conclusion}

In this paper, we proposed a novel model named SePEnTra for TEM that integrates security and privacy during energy trading. SePEnTra has five phases and uses two well-known cryptographic techniques; additive secret sharing and Pedersen commitment. By employing additive secret sharing and Pedersen commitment, the market user can share energy forecasts and actual data with the market operator and other users without violating individual’s privacy while the market operator utilizes this information to determine honest price signals in the forecast phase and detect malicious users’s activity in the online phase.   

As a system model, we consider a community-based P2P market with distributed pricing mechanism in this work. We have analyzed the security of our proposed model in detail. Then, we theoretically and numerically evaluate the performance of our model, taking into account large and practical scenarios. Finally, we have compared the performance of our model with the same TEM framework without considering security measures. The results show that the proposed framework's computation complexity and communication overhead are low, and it is storage-efficient while providing security and privacy by using two advanced cryptographic techniques. This work can be extended in the future by implementing SePEnTra in a hybrid P2P market.

\bibliographystyle{IEEEtran}
\bibliography{main}

\vspace{-2cm}

\begin{IEEEbiography}[{\includegraphics[width=1in,height=1.25in,clip,keepaspectratio]{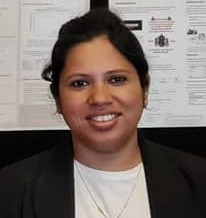}}]{Rumpa Dasgupta}
received a B.S. degree in computer science and engineering from the Chittagong University of Engineering and Technology (CUET), Bangladesh, in 2010 and the M.S. degree from the University of Ulsan, South Korea, in 2017. She is currently a 3rd year Ph.D. student at the Faculty of IT, Monash University. Her research interests include privacy and security issues of cyber-physical systems.
\end{IEEEbiography}
\vskip -2\baselineskip plus -1fil 
\begin{IEEEbiography}
[{\includegraphics[width=1in,height=1.25in,clip,keepaspectratio]{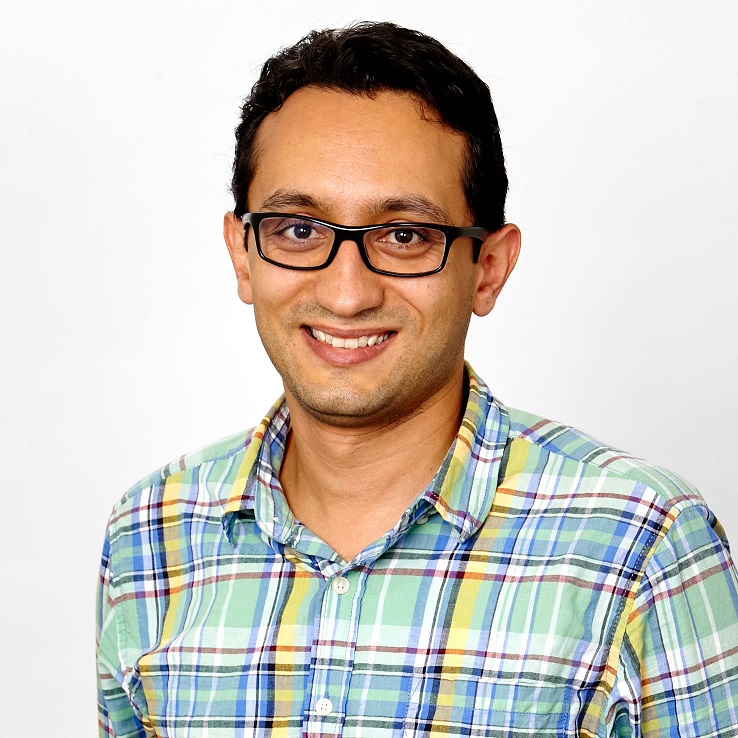}}]{Amin Sakzad}
received his Ph.D. in applied
Mathematics from the Amirkabir University of Technology (AUT), Tehran, Iran, 2011. He is a Senior Lecturer at the Faculty of IT, Monash University, Melbourne, Australia. He is mainly interested in the applications
of Euclidean lattices in cryptography and wireless communications. 
\end{IEEEbiography}
\vskip -2\baselineskip plus -1fil 
\begin{IEEEbiography}
[{\includegraphics[width=1in,height=1.25in,clip,keepaspectratio]{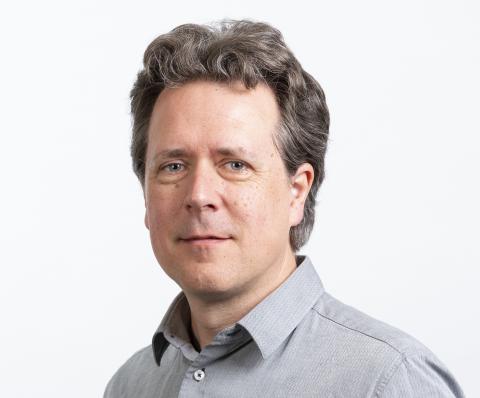}}]{Carsten Rudolph}
is a Professor of the Faculty of IT at Monash University and Director of the Oceania Cyber Security Centre OCSC. Carsten Rudolph has contributed extensively to
four key areas of cybersecurity: Trusted Computing, Security of critical infrastructures and Security of IT Networks; Security by design; formal methods for security; Validation and design of security protocols; Digital
forensic readiness and secure digital evidence.
\end{IEEEbiography}
\vskip -2\baselineskip plus -1fil 
\begin{IEEEbiography}
[{\includegraphics[width=1in,height=1.25in,clip,keepaspectratio]{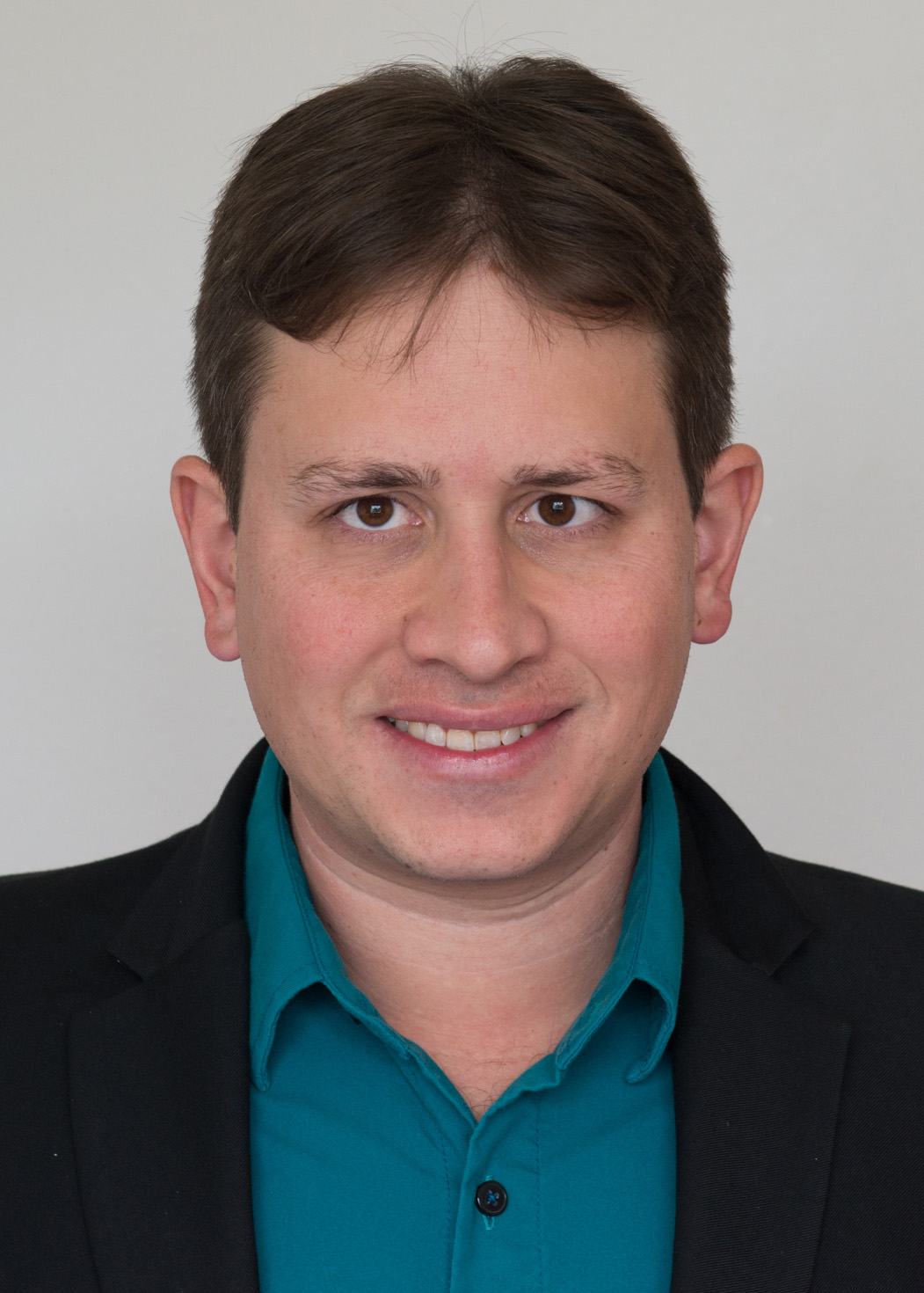}}]{Rafael Dowsley}
received the Ph.D. degree from KarlsruheInstitute of Technology, Germany, 2016. He is a Lecturer of the Faculty of Information Technology, Monash University, Melbourne, Australia. His research interests lie in cryptography and its many intersections with other fields such
information theory, machine learning, privacy and security. 
\end{IEEEbiography}
\end{sloppypar}

\end{document}